\documentclass[12pt]{iopart}
\usepackage{geometry}                
\usepackage{graphicx}
\usepackage{amssymb}
\usepackage{dsfont}
\usepackage{epstopdf}
\usepackage{color}
\usepackage{mathtools}
\usepackage{cancel}
\usepackage{amsmath}

\definecolor{red}{rgb}{1,0,0}
\definecolor{blue}{rgb}{0,0,1}
\definecolor{black}{rgb}{0,0,0}

\newcommand{\p}{\partial}
\newcommand{\eq}[1]{\begin{align}#1\end{align}}
\newcommand{\eqs}[1]{\begin{align*}#1\end{align*}}
\newcommand{\ffrac}[2]{\mbox{$\frac{#1}{#2}$}}
\newcommand{\half}{\mbox{$\frac{1}{2}$}}
\newcommand{\hhalf}{\frac{1}{2}}
\newcommand{\OO}{\mathcal{O}}
\newcommand{\D}{{\mathcal{D}}}

\renewcommand{\S}{{\mathcal{S}}}

\newcommand{\ov}{\vec{1}}
\newcommand{\eh}{\hat{e}}
\renewcommand{\a}{\alpha}

\usepackage{calrsfs}
\newcommand{\co}{c^{\circ}}

\newcommand{\phisv}{\vec{\phi}^*}
\newcommand{\phiv}{\vec{\phi}}
\newcommand{\muv}{\vec{\mu}}
\newcommand{\nuv}{\vec{\nu}}

\newcommand{\phis}{\phi^*}

\newcommand{\pv}{\vec{p}}
\newcommand{\qv}{\vec{q}}

\newcommand{\xiv}{\vec{\xi \;}}

\newcommand{\mv}{\vec{m}}
\newcommand{\nv}{\vec{n}}
\newcommand{\sv}{\vec{s}}
\newcommand{\zv}{\vec{z}}
\newcommand{\Gv}{\vec{G}}

\newcommand{\ah}{\hat{a}}
\newcommand{\ahd}{\hat{a}^\dagger}
\newcommand{\on}{\overline{n}}

\definecolor{darkblue}{rgb}{0,0,0.6}
\usepackage[colorlinks,linkcolor=darkblue,citecolor=darkblue,urlcolor=darkblue]{hyperref}

\begin{document}

\title[Dynamical mean-field theory: from ecosystems to reaction networks]{Dynamical mean-field theory: from ecosystems to reaction networks}

\vspace{.5cm}

\author{Eric De Giuli$^1$ and Camille Scalliet$^2$}
\address{$^1$ Department of Physics, Toronto Metropolitan University\footnote{formerly Ryerson University}, M5B 2K3, Toronto, Canada}
\address{$^2$ Department of Applied Mathematics and Theoretical Physics, University of Cambridge, Wilberforce Road, Cambridge, CB3 0WA, United Kingdom} 
\ead{edegiuli@ryerson.ca,cs2057@cam.ac.uk}
\vspace{.5cm}
\begin{indented}
\item[]November 2022
\end{indented}      

\vspace{1cm}

\begin{abstract}
Both natural ecosystems and biochemical reaction networks involve populations of heterogeneous agents whose cooperative and competitive interactions lead to a rich dynamics of species' abundances, albeit at vastly different scales. The maintenance of diversity in large ecosystems is a longstanding puzzle, towards which recent progress has been made by the derivation of dynamical mean-field theories of random models. In particular, it has recently been shown that these random models have a chaotic phase in which abundances display wild fluctuations. When modest spatial structure is included, these fluctuations are stabilized and diversity is maintained. If and how these phenomena have parallels in biochemical reaction networks is currently unknown. Making this connection is of interest since life requires cooperation among a large number of molecular species. In this work, we find a reaction network whose large-scale behavior recovers the random Lotka-Volterra model recently considered in theoretical ecology. We clarify the assumptions necessary to derive its large-scale description, and reveal the underlying assumptions made on the noise to recover previous dynamical mean-field theories. Then, we show how local detailed balance and the positivity of reaction rates, which are key physical requirements of chemical reaction networks, provide obstructions towards the construction of an associated dynamical mean-field theory of biochemical reaction networks. Finally, we outline prospects and challenges for the future. \end{abstract}
\begin{indented}
\item[]{\it Keywords}: Disordered systems, Statistical field theory, Theoretical ecology, Reaction networks.
\vspace{-.5cm}
\item[] \submitto{\jpa}
\end{indented}     
\maketitle

\section{Introduction}

The development of a theoretical framework to describe life as a physical system, with relevant order parameters that distinguish animate from inanimate matter, is a major goal of 21st century physics \cite{Laughlin00,Goldenfeld11}. First, it has long been accepted that such a framework should deal with conditions far from thermodynamic equilibrium \cite{schrodinger}. The absence of a Boltzmann distribution forces one to generally consider the full dynamics of out-of-equilibrium systems. Ideally, the theoretical framework should be dynamic in nature. A second key aspect of life is that it involves, at any scale, the coordinated action of a large number $N$ of chemical species, which interact via a complex set of chemical reactions. This opens the possibility of exploiting a large-species $N \to \infty$ limit to obtain a tractable theory, with the prospect of obtaining universal results. This presence of variability at the level of species chemical interactions also adds motivation to consider simplified models with random interactions.

Dynamical mean-field theory (DMFT) is essentially the unique framework that combines both of the above essential aspects of the living state. It is a theoretical approach to disordered statistical models that starts from a mesoscopic description. The disorder is typically introduced via the couplings between degrees of freedom. The method relies on the assumption that the number of degrees of freedom in the system is large; formally the thermodynamic limit of an infinite number of species is taken. This dynamic mean-field theory describes the time evolution of a typical degree of freedom after the average over the quenched disorder has been carried out, 
and yields a closed description of the dynamics in terms of an effective species and a small number of order parameters. First developed to describe the dynamics of spin-glasses \cite{Sompolinsky82}, DMFT has since been successful in identifying and describing phase transitions in other areas of physics, such as neural networks \cite{Sompolinsky88,Kadmon15}, structural glasses \cite{Maimbourg16,szamel17,Agoritsas19}, and ecosystems \cite{Opper92,Pearce19,Roy20}. Dynamical mean-field theory thus appears as a promising framework to describe life as a physical system characterized by distinct dynamical phases.

Recently, substantial progress was made to obtain a dynamical mean-field theory of ecosystems, using the generalized Lotka-Volterra model as a starting point \cite{Opper92,Pearce19,Roy20}. This theory has shed light on the `diversity paradox', namely the counterintuitive fact that many ecosystems are much more diverse than what is expected from niche-filling.  Importantly, the theory has revealed that dynamical models for ecosystems can exist in a variety of phases, characterized either by a single equilibrium, multiple marginally-stable equilibria, or chaos \cite{Bunin17,Biroli18}. The framework has thus allowed extension of the notion of phase away from thermodynamic equilibrium.

Compared to ecosystems, the organization of life occurs primarily at much smaller scales, where molecules are the main actors. As stressed by Anderson~\cite{anderson1983suggested}, life requires a diversity of chemical species, which are often involved in complex chemical reactions networks (CRN). Even in the constant temperature case common in biology, CRNs can exhibit rich dynamical behavior: unstable steady states, multiple steady states, sustained composition oscillations, and chaotic dynamics \cite{craciun2006multiple,bray1921periodic,briggs1973oscillating}. These behaviors have a parallel in model ecosystems, where their existence has been firmly established by DMFT. 

Our article builds naturally on past works employing a reaction network representation of ecological models \cite{Butler09,Shih14,Xue17,Constable16,Constable17}. Our interest is however not only to use CRN as a means to study model ecosystems, but to investigate to what extent the recent advances made possible by the development of DMFT in theoretical ecology can advance our understanding of emergent behavior in CRNs. We specifically investigate the behavior of CRNs in the limit of many chemical species which, to the best of our knowledge, has not been considered in the past.

The overarching goal of this article is to ask: to what extent can one develop a DMFT for general chemical reaction networks? Our strategy is to first find a chemical reaction network whose large-scale behavior yields the generalized Lotka-Volterra (GLV) model studied in ecology. We then examine which features of such a GLV reaction network are universal, and which features of general reaction networks are absent from the mapping. We will conclude that the present DMFTs for ecology do not apply to generic chemical reaction networks, suggesting the existence of universality classes of DMFTs yet to be explored.

The article is organized as follows. In section \ref{background}, we review the two existing analytical techniques on which our work is based: dynamical mean field theory, and the Doi-Peliti path integral formalism for chemical reaction networks. In section \ref{GLV} we find a microscopic reaction network which maps to the generalized Lotka-Volterra model studied in the ecology literature. We show that the derivation of the ecology model is not systematic and requires stringent, and in general unjustified, assumptions. We then derive a generalized dynamical mean-field theory starting from the path integral, and recover the large-species DMFT obtained previously. In section \ref{mapping}, we explore the consequences of the mapping between chemical reaction networks and theoretical ecology. In section \ref{discussion}, we conclude by discussing future directions for the development of new classes of DMFT.

\section{Background}
\label{background}

In this section, we review two existing analytical techniques that we will combine in this article: (i) dynamical mean-field theory, and (ii) the Doi-Peliti path integral formulation for chemical reaction networks. 

\subsection{A brief review of dynamical mean-field theory}  

Dynamical mean-field theory (DMFT) is a general technique which allows to describe the dynamical behaviour of mean-field disordered systems with a minimal number of relevant order parameters. It is distinct from the dynamical mean-field theory of correlated electron systems \cite{Georges96}, which describes the behaviour of a quantum-mechanical system in terms of an effective impurity. DMFT was first introduced to study the dynamics of mean-field spin glasses \cite{Sompolinsky82}, and later applied to the spherical $p$-spin model \cite{Kirkpatrick87,Cugliandolo93}. The long-time asymptotic solution to the equations for the latter exhibits the aging phenomenon and explains it in terms of an effective temperature, a great achievement for DMFT \cite{Cugliandolo93}. Since then, DMFT has successfully been applied to neural networks \cite{Sompolinsky88,Kadmon15}, structural glasses \cite{Maimbourg16,szamel17,Agoritsas19}, and ecosystems \cite{Opper92,Pearce19,Roy20}. To the best of our knowledge, DMFT has not previously been applied to chemical reaction networks.

The main ingredients to construct a dynamical-mean field theory are the following. Consider a generic theory with dynamical variables $\sigma_i(t)$ with $i = 1,\ldots,N$ in the presence of disorder. Depending on the context, these could be spin variables, or species abundances, for which disorder is introduced via random coupling coefficients, or random inter-species interactions. Dynamical mean-field theories generally begin with a Langevin dynamics for the variables $\sigma_i(t)$. Either by path-integral methods \cite{DeDominicis78,msr,Sompolinsky82}, or the dynamical cavity method~\cite{mezard1987spin}, one performs a disorder average to eliminate it. The many-species dynamical problem is then reduced to an effective process for a representative degree of freedom. A consequence of the disorder average is however to introduce two-time quantities. The most basic of these are the correlation function $C(t,t') = \ffrac{1}{N} \sum_i \langle \sigma_i(t) \sigma_i(t') \rangle$ and the response function $R(t,t') = \ffrac{1}{N} \sum_i \langle \left. \delta \sigma_i(t)/ \delta h_i(t') \right|_{h=0} \rangle$, where $h_i(t)$ is a field conjugate to the dynamical variable $\sigma_i(t)$. Here $\langle \cdot \rangle$ denotes a thermal average subject to appropriate initial conditions. In the simplest class of DMFTs, one obtains a pair of integro-differential equations involving only $C(t,t')$ and $R(t,t')$ \cite{Sompolinsky82,Kirkpatrick87}. One can also find theories for which the equations for the evolution of $C$ and $R$ involve a memory kernel that must be determined from a self-consistent 1D stochastic process \cite{Opper92,Maimbourg16,Pearce19,Roy20}. More generally, we include in the DMFT category any theory that reduces the number of degrees of freedom from $N \to \infty$ down to a finite number. We will distinguish multiple levels of complexity in our work. 

\subsection{A brief review of the Hamiltonian formulation of chemical reaction networks}  

We consider a closed volume $\Omega$ containing a mixture of $N$ chemical species denoted $X_i$ with $i=1,...,N$. We note $n_i(t)$ the number of molecules $X_i$ at time $t$. We focus on the well-mixed limit and consider homogeneous concentrations, expressed in units of $\co$, the standard concentration 1 mol/L. The state of the mixture is fully specified by the vector $\nv(t)$ of the number of molecules of each species, which changes when a chemical reaction occurs. We consider chemical reactions, labelled by $\alpha$, of the form 
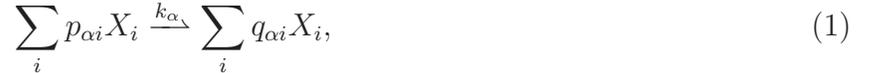
\eq{ \label{eq1}
\sum_i p_{\alpha i} X_i \xrightharpoonup{k_\alpha} \sum_i q_{\alpha i} X_i,
}
where $k_\alpha$ is the rate of reaction $\alpha$. By convention, $p_{\alpha i}$ and $q_{\alpha i}$ are the stoichiometric coefficients for the molecular species $X_i$ as reactant and product, respectively. The complete set of reactions is understood to contain each reaction in both forward and backward directions. 
In full generality, reactions need not conserve the number of molecules, thus allowing interactions with external reservoirs.

\subsubsection{Chemical master equation}

While the rate equations describe the time evolution of the average number of molecules, they do not capture the fluctuations intrinsic to the stochastic nature of chemical reactions. To obtain a complete and closed description of the dynamics, one must track the evolution of the probability distribution $P(\nv,t)$, as described in Ref.~\cite{Van-Kampen92}. This yields the chemical master equation
\eq{ \label{master}
\frac{\p P(\nv,t)}{\p t} = \sum_\alpha \tilde k_\alpha  \left(\prod_i \mathcal{E}_i^{p_{\alpha i}-q_{\alpha i}} - 1 \right) \prod_j \left(\frac{n_j!}{(n_j-p_{\alpha j})! } \right) P(\nv,t),
}
where the products run over the $N$ species, and $p_{\alpha j}=0$ if $X_j$ does not intervene in reaction $\alpha$. The probability for a collision involving $p_{\alpha j}$ molecules $X_j$ is proportional to $n_j(n_j-1)...(n_j-p_{\alpha j}+1)/(\co \Omega)^{p_{\alpha j}}$. This gives rise to the factorial terms in equation \eqref{master}, and the rescaled reaction rate $\tilde k_\alpha = k_\alpha (\co\Omega)^{1-\sum_i p_{\alpha i}}$. The effect of each reaction $\alpha$ is to transform $n_j$ into $n_j + q_{\alpha j} - p_{\alpha j}$. This is conveniently captured by the step operator $\mathcal{E}_i$ which acts on any arbitrary function $f(\nv)$ as
\eqs{
\mathcal{E}_i f(\vec{n}) = f(\nv + \vec{1}^i), \quad \mathcal{E}^{-1}_i f(\vec{n}) = f(\nv - \vec{1}^i),}
with $1^i_j = \delta_{ij}$. We assume a well-stirred setting in which there is no spatial dependence of concentration, although the derivation straightforwardly extends to the more general case \cite{Doi76,Peliti85,Tauber14}. 

However, the chemical master equation \eqref{master} cannot be solved analytically for generic chemical reaction networks. 

\subsubsection{Doi formalism}

To make progress, we adopt the Hamiltonian formalism developed by Doi \cite{Doi76} which consists in an exact rewriting of the chemical master equation \eqref{master}. Doi observed that the step operators in \eqref{master} and associated factorial terms $n_j!/(n_j-p_{\alpha j})!$ have a natural representation in terms of bosonic creation and annihilation operators. Reaction systems thus have a natural interpretation in terms of quantum mechanics through second quantization. 

In this formalism, one introduces a non-Hermitian quasi-Hamiltonian, or Liouvillian, $H$. Unlike the rate equations, the Hamiltonian $H$ completely specifies the reaction network, as the chemical master equation does. Doi's construction and the corresponding path integral formulation are reviewed in \ref{appendixDP}. The resulting Hamiltonian $H$ is a sum of contributions from each reaction, $H = \sum_\alpha H_\alpha$, with 
\eq{ \label{H1}
H_\alpha(\phiv,\phisv) & = \tilde k_\alpha \left[ \prod_j \phis_j{}^{q_{\alpha j}} - \prod_j \phis_j{}^{p_{\alpha j}} \right] \prod_i \phi_i^{p_{\alpha i}}.
}
The arguments $\phiv,\phisv$ are abstract quantities that get introduced first via creation and annihilation operators, and then as complex variables in the path-integral construction. Only their product, $\phis_j \phi_j = n_j$, has a direct physical interpretation as the number of molecules of $X_j$. We thus perform a Hopf-Cole transformation \cite{Kamenev02,Andreanov06,Lefevre07,Smith11,Smith20} 
\eq{ \label{hopfcole}
\phis_j=e^{\nu_j}, \quad \phi_j = n_j e^{-\nu_j}, 
}
in terms of which we have
\eq{ \label{H2}
H_\alpha(\nv,\nuv) & = \left[ e^{\nuv \cdot \vec{s}_\alpha } - 1 \right] \underbrace{\tilde k_\alpha  \prod_i n_i^{p_{\alpha i}}}_{\equiv F_\alpha(\nv)} ,
}
where we introduced the stoichiometric matrix ${s}_{\alpha j} = q_{\alpha j} - p_{\alpha j}$. Here we are abusing notation slightly to write $H$ as a function of its new arguments $\nv$ and $\nuv$, in place of the previous ones. In what follows we always intend this form for $H$. The doubling of degrees-of-freedom in Eqs.\eqref{H1},\eqref{H2} is characteristic of dynamic problems \cite{De-Dominicis76,Janssen76,Kamenev02}.

The $\nuv$ variables do not have a straightforward interpretation. They can be considered as a per-species bias or `tilt' \cite{Smith20}. As explained below, $\nuv = 0$ corresponds to mean-field trajectories.

Several features of \eqref{H2} are important. First, the $\nuv$-dependent factor is independent of the kinetics: conservation of probability requires that $H_\alpha(\nv,0) = 0$ for any $\nv$, and the specific form in \eqref{H2} follows from the Doi-Peliti construction for reaction networks with a stoichiometric matrix ${s}_{\alpha j}$. Second, the kinetics are encoded in the function $F_\alpha(\nv)$, which can be written as 
\eq{ \label{F}
F_\alpha(\nv) = k_\alpha (\co\Omega) \prod_i \left( \frac{n_i}{\co \Omega} \right)^{p_{\alpha i}}.
}
The specific form in \eqref{F} corresponds to the usual mass-action kinetics, but other choices are possible, so long as $F_\alpha(\nv) \geq 0$. Finally, the Hamiltonian \eqref{H2} specifies only the dynamics of the system, and does not contain any information about thermodynamics. We will return to the consequences of thermodynamics, in particular the detailed balance condition, below. 

\subsubsection{Peliti integral}

Using Doi's Hamiltonian formulation, Peliti proposed a path-integral description of reaction systems, constructed as in quantum mechanics \cite{Peliti85}. In principle, one can obtain the complete statistics of species abundances by evaluating the path integral
\eqs{ 
Z = \int \D n \int \D\nu \; e^{-S},
}
with the action
\eq{ \label{S1}
S = \int_0^{t_f} dt \left[ \nuv \cdot \p_{t} \nv - H(\nv,\nuv) \right] + S_{BC},
}
where 
\eqs{
S_{BC} =  [\nv(0)-\nv^0] \cdot \nuv(0) + \sum_j n_j(t_f) [-e^{-\nu_j(t_f)} + 1 - \nu_j(t_f) ],
}
encodes the boundary conditions at initial $t=0$, and final $t_f$ times. Here we consider $n_i(0) = n^0_i$ as initial condition.

In practice, observables are computed by adding sources, or fields, to the partition function $Z$. For that purpose, we consider the generating function 
\eqs{
Z(\zv,t_f) = \sum_{\{\nv\}} z_1^{n_1} z_2^{n_2} \cdots z_N^{n_N} P(\nv,t_f).
}
The partition function is recovered for $\zv=1$, and moments are computed by derivatives at $\zv=1$. Elgart and Kamenev \cite{Elgart04} showed that values $\zv$ different from unity are necessary to understand rare fluctuations. They also showed that the path-integral representation of the generating function is that of the original partition function ($\zv=1$), equation \eqref{S1}, with differences only in the boundary terms
\eqs{
S_{BC} = \sum_j \left[ [\nv(0)-\nv^0] \cdot \nuv(0) + n_j(t_f) [-z_j e^{-\nu_j(t_f)} + 1 - \nu_j(t_f) ] \right].
}
This gives boundary conditions $\nuv(t_f) = \log \zv$, $\nv(0) = \nv^0$. The particle number statistics are extracted from the generating function by appropriate contour integrals. For example, the marginal distribution of the $j^{th}$ species is
\eqs{
\rho_j(m_j) & = \oint \frac{dz_j}{2\pi i} \frac{1}{z_j^{m_j+1}} Z(\ov + (z_j-1) \eh_j,t)
}
where $\ov = (1,1,\ldots,1)$ and $\eh_j = (0,\ldots,0,1,0,\ldots,0)$ with a 1 in the $j^{th}$ position, and the integral is around the unit circle. Physical quantities are thus determined by $Z(\zv,t_f)$ in the vicinity of $\zv=\ov$. 

To evaluate the path integral $Z$ in practice, one resorts to approximations. Since $\nuv(t_f) = \log \zv$, and we need to consider $Z$ in the vicinity of $\zv=\ov$, this motivates a small-$\nu$ expansion. From conservation of probability, the action begins with $\OO(\nu)$. To leading order, $\OO(\nu)$ in $S$, one obtains the classical rate equations. To the next order, $\OO(\nu^2)$, one obtains the chemical Langevin equation, derived below. 

\subsubsection{Instanton equations} Instead of making a small-$\nu$ expansion of the action, one can look for saddle-points, i.e. paths that give the largest contribution to the integral. To motivate this, we consider a naive scaling analysis of the action. Suppose that the species numbers are large, $n \gg 1$. Since the number of reactants $p_{\alpha i}$ is generally varying from reaction to reaction, we see from \eqref{H2} and \eqref{F} that in order to have nontrivial dynamics for general reaction networks,  we should have $\co \Omega \gg 1$, with the dimensionless concentrations $n_i/(\co \Omega) = \OO(1)$. In this case $H \sim \co \Omega \gg 1$ and a saddle-point analysis of the path integral is formally justified. Following the nomenclature of field theory we call these saddle-point paths `instantons' \cite{Coleman88}. The instanton equations, derived in \ref{appendixinstanton}, are

\begin{subequations} \label{inst} \eq{ 
\frac{\p \nv}{\p t} & = +\frac{\p H}{\p \nuv} = \sum_\alpha F_\alpha(\nv)  \sv_\alpha e^{ \nuv \cdot  \sv_\alpha } \label{inst1}, \\
\frac{\p \nuv}{\p t} & = -\frac{\p H}{\p \nv} = \sum_\alpha \frac{\p F_\alpha(\nv)}{\p \nv} \left[ e^{\nuv \cdot \sv_\alpha } - 1 \right].  \label{inst2}
} \end{subequations}
We see immediately that $\nuv = 0$, corresponding to mean-field trajectories, is always a solution. In that case \eqref{inst1} reduces to the reaction rate equations. 

Eqs.\eqref{inst1},\eqref{inst2} take the form of Hamilton's equations, so several properties familiar from classical mechanics also hold here. In particular, along an instanton the time-evolution of the Hamiltonian is
\eqs{
\frac{dH}{dt}  = \frac{\p H}{\p \nuv} \cdot \frac{\p \nuv}{\p t} + \frac{\p H}{\p \nv} \cdot \frac{\p \nv}{\p t} + \frac{\p H}{\p t} = \frac{\p H}{\p t}.
}
We see that if the Hamiltonian is time-independent, it is conserved by the dynamics along the instantons.

\subsubsection{Expansion to obtain the Langevin equation}
\label{langevin}
As explained in \ref{appendixlangevin}, keeping terms to order $\OO(\nu^2)$ in the action, one obtains the chemical Langevin equation
\eq{ \label{Lan1}
\p_{t} \nv = \vec{h}(\nv,t) + B(\nv,t)^{1/2} \cdot \xiv,
}
where the drift $\vec{h}$ and correlation matrix $B$ are related to the Hamiltonian
\eq{
\vec{h}(\nv,t) &= \p_{\nuv} H(\nv(t),0) \label{Lan2}, \\
 B(\nv,t) &= \p_{\nuv} \p_{\nuv} H(\nv(t),0) \label{Lan3},
 }
and $\xiv$ is a white noise
\eqs{ 
\langle \xi_j(t) \rangle & = 0, \quad \langle \xi_j(t) \xi_k(t') \rangle = \delta(t-t') \delta_{jk}.
} 
The boundary conditions are $\nv(0) = \nv^0$ and trajectories are weighted with $e^{-S}$, with
\eqs{
S = \half \sum_j \frac{n_j(t_f) (1-z_j^2)}{z_j} + \half \sum_j \log (z_j n_j(t_f)).
}
The late-time boundary contribution is specific to the generating function; alternative boundary conditions are discussed in the \ref{appendixlangevin}. 

From the derivation of the Langevin equation we can infer that $z-1 \sim 1/\sqrt{n}$, thus $\nu \sim \log z \sim 1/\sqrt{n}$. This justifies {\it a posteriori} the small-$\nu$ expansion for large systems, and is also consistent with the saddle-point analysis, since the action scales as $n/\sqrt{n}\gg 1$. More precisely, for a species $j$ the small-$\nu_j$ expansion is justified when $n_j$ is large. This small-$\nu$ expansion is thus the system-size expansion in path-integral language. 

\subsubsection{Thermodynamics and detailed balance}
To this point, we have only considered the kinetics of reaction networks. We now turn to the constraints imposed by thermodynamics. Consistency with thermodynamics requires that if a reaction $\alpha$ occurs with rate $k^+_\alpha$, then its reverse must occur with rate $k^-_\alpha$. The ratio of the rates is fixed by $(\Delta G)_{\alpha} = (\qv_{\alpha} - \pv_{\alpha}) \cdot \Gv$, the difference in molar Gibbs free energy between products and reactants, expressed in units of $RT$
\eq{ \label{TST1}
\frac{k^+_\alpha}{k^-_\alpha} = e^{-(\Delta G)_\alpha},
} 
which is known as local detailed balance \cite{Van-den-Broeck13}. Using modern transition state theory \cite{Gilbert90}, one can predict from first principles 
\eq{ \label{TST2}
k^+_\alpha = k_0 \; e^{-(\delta G)_\alpha},
}
where we introduced $(\delta G)_\alpha = G_{A_\alpha} - \pv_{\alpha} \cdot \Gv$, the difference in molar Gibbs free energy between the activated complex $A_\a$ of reaction $\alpha$ and the reactants, and $k_0 = 1/(2\pi \beta \hbar) \approx 6 \times 10^{12}$ Hz at 300K. Eq.~\eqref{TST2} holds for both forward and backward reactions, {\it mutatis mutandis} so that \eqref{TST1} holds
\eqs{
\frac{k^+_\alpha}{k^-_\alpha}  = e^{-(G_{A_\alpha} - \sum_i p_{\alpha i} G_i)} e^{+(G_{A_\alpha} - \sum_i q_{\alpha i} G_i)_\alpha} = e^{-(\Delta G)_\alpha}.
}
This allows us to rewrite the kinetic part of the Hamiltonian as
\eq{ \label{F2}
F_\alpha(\nv) = k e^{-G_{A_\alpha}} \prod_i \left( \frac{n_i}{c^{eq}_i \Omega} \right)^{p_{\alpha i}},
}
with $k= k_0 \co \Omega$ and $c^{eq}_i = \co e^{-G_i}$.

When a reaction occurs in a forward-reverse pair $\alpha_+$ and $\alpha_-$, its contribution to the right-hand side of the $\p n/\p t$ instanton equation is
\eqs{
F_{\alpha_+} \sv_{\alpha_+} e^{\nuv \cdot \sv_{\alpha_+}} + F_{\alpha_-} \sv_{\alpha_-} e^{\nuv \cdot \sv_{\alpha_-}} = \sv_{\alpha_+} k e^{-G_{A_\alpha}} \left[ e^{\nuv \cdot \sv_{\alpha_+}} \prod_i \left( \frac{n_i}{c^{eq}_i \Omega} \right)^{p_{\alpha i}} - e^{\nuv \cdot \sv_{\alpha_-}} \prod_i \left( \frac{n_i}{c^{eq}_i \Omega} \right)^{q_{\alpha i}}  \right]
}
The bracket vanishes if $n_i = c^{eq}_i \Omega$ and $\nu_i=0$ for all $i$. Thus if \textit{all} reactions occur in such pairs, this is a steady-state solution. This is the well-known expression for thermal equilibrium in a CRN \cite{Van-Kampen92}. 

The fact that chemical reactions satisfy detailed balance has profound physical and conceptual consequences. In particular, this makes the system time-reversal symmetric, at least statistically. For any field theory, symmetries of the system appear as symmetries in the action. There is thus a symmetry operation for the action that reverses time, as detailed in \ref{appendixtrs}. When all reactions occur in forward-reverse pairs satisfying local detailed balance, the transformation
\eqs{ 
\tilde \nu_i(t_f-t) & = -\nu_i(t) + \log (n_i(t)/(c^{eq}_i \Omega)), \\
\tilde n_i(t_f-t) & = n_i(t),
}
is a symmetry of the action. Note that the transformation is nonlocal in time. Steady solutions, that are invariant under time-reversal, satisfy $2\nu_i = \log (n_i/(c^{eq}_i \Omega))$. The thermal equilibrium solution $n_i/\Omega = c^{eq}_i$ is recovered when $\nu_i=0$, as expected \cite{Van-Kampen92}. 

\subsubsection{Existence of Lyapunov function}
\label{lyapunov}

We have shown that $\nuv = 0$ along mean-field solutions, hence $H=0$ from equation (\refeq{H2}). Since $H$ is conserved by the instanton dynamics, we also have $H=0$ on relaxations toward mean-field solutions. The locus of points defined by $H=0$ thus plays a special role \cite{Elgart04,Kamenev02,Smith20}. Let us implicitly define a functional $\Psi(\nv)$ by considering the equation $H[\muv, \nv] = 0$ where $\muv = \p \Psi/\p \nv$. This is a time-independent Hamilton-Jacobi equation \cite{Smith20,Hong20}. Let us compute the derivative of $\Psi$ along the deterministic dynamics $\nuv=0$ \cite{Hong20}
\eqs{
\frac{d \Psi}{dt} = \left. \frac{\p \Psi}{\p \nv} \cdot \frac{d \nv}{dt}  \right|_{\nuv=0}  = \left. \muv \cdot \frac{\p H }{\p \nuv}  \right|_{\nuv=0} = \muv \cdot \sum_\alpha F_\alpha(\nv) \sv_\alpha,
}where we have used Hamilton's equation (\refeq{inst1}). Subtracting $H[\muv,\nv]=0$ (\refeq{H2}), we find that $\Psi$ is non-increasing under the deterministic flow
\eqs{ 
\frac{d \Psi}{dt} & = \sum_\alpha F_\alpha(\nv) \left[ \muv \cdot  \sv_{\alpha}  - e^{\muv \cdot \sv_\alpha} + 1 \right] \leq 0,
}
since $e^a- 1 - a \geq 0$ for all real $a$, and $F_\alpha \geq 0$. We have shown that $\Psi$ is a Lyapunov function for the deterministic dynamics. The quantity $-\Psi$ has the interpretation of a non-equilibrium entropy functional, and $-d\Psi/dt$ corresponds to the total entropy change, including both entropy production and entropy exchanges \cite{Hong20}. 

The existence of a Lyapunov function for chemical reaction networks in the deterministic limit has important consequences. In fact, complex chemical reaction networks may have multiple attractors \cite{Fang19,Qian16,Wang11,Smith11,Smith20}. In the limit of small systems, hopping between these attractors may occur, giving rise to interesting dynamical behavior. The existence of a Lyapunov function in the deterministic limit, recovered for large system size, however prevents such hopping dynamics. 

A Lyapunov function is also a practical tool, since an explicit expression for it can be used to study the statistics of attractors. When noise is present, then at the level of the Langevin dynamics, there may be no Lyapunov function, but still there is an effective Hamiltonian governing the long-time behaviour \cite{Parisi88}. For the Lotka-Volterra model studied below, such an effective Hamiltonian has been used ~\cite{Biroli18}.

\section{Generalized Lotka-Volterra model as a reaction network}
\label{GLV}

\subsection{Generalized Lotka-Volterra model} Consider an isolated ecosystem with $N$ species $i=1,\ldots,N$ and abundances $n_i(t)$ of each species. We assume that the ecosystem is well-mixed and do not consider spatial dependences of the abundances. A minimal model of population dynamics in the ecosystem must include a birth rate and death rate for each species, along with inter-species interactions. The generalized Lotka-Volterra (GLV) model includes these main features. It also includes a small immigration rate $\lambda$ which prevents extinctions. The stochastic nature of births and deaths, causing populations to fluctuate even without external noise, is captured within the GLV by demographic noise. We are eventually interested in large ecosystems, whose complexity and inherent randomness is modeled by taking random inter-species interaction strengths \cite{may72}. We consider the GLV model in a Langevin form
\eq{ \label{LV1}
\frac{dn_i}{dt} = \frac{r_i}{\kappa_i} (n_i \kappa_i - n_i^2) - \sum_{j\neq i} S_{ij} n_i n_j + \lambda + \sqrt{n_i} \xi_i,
}
where $r_i$ and $\kappa_i$ are respectively the birth rate and carrying capacity of species $i$,  and $\xi_i$ is a white noise with $\langle \xi_i \rangle = 0, \langle \xi_i(t) \xi_j(t') \rangle = 2 \omega^2 \delta_{ij} \delta(t-t')$. The interaction matrix elements $S_{ij}$ are Gaussian random variables, with mean, variance and asymmetry given by
\eq{
\overline{S_{ij}}=s/N, \quad \overline{(S_{ij}-\overline{S_{ij}})^2} = \sigma^2/N, \quad \overline{(S_{ij}-\overline{S_{ij}})(S_{ji}-\overline{S_{ij}})} = \gamma \sigma^2/N,\label{PS}
}with a scaling of the cumulants ensuring a proper large ecosystem limit $N \gg 1$. The parameter $-1 \leq \gamma \leq 1$ characterizes the degree of asymmetry in the inter-species interactions.

The generalized Lotka-Volterra model Eq.~\eqref{LV1} has been studied in the limit of many interacting species $N\gg 1$, both in the deterministic limit where $\omega=0$ \cite{Bunin17,Barbier18} for general $r_i$ and $\kappa_i$, and with finite demographic noise \cite{Altieri20} with parameters $r_i=1, \kappa_i=1$. The phase diagram was also established with and without demographic noise in the limit of symmetric strongly competitive interactions \cite{Altieri2022}. The species-local part of the right-hand side ($n_i \kappa_i - n_i^2$) can be generalized to other polynomials $f_i(n_i)$, see \cite{Biroli18,Altieri20}. Recently, a DMFT has also been built for the GLV considering structured, non-random, interactions \cite{poley2022generalised}.

Although one might expect demographic noise to be irrelevant in large populations, it is crucial when the underlying system has multiple attractors, as discussed below. Moreover in related models, its importance has been emphasized in its effect on spatial patterning \cite{Durrett94,Butler09}, temporal oscillations \cite{Shih14}, ecology-evolution interactions \cite{Xue17}, and on the full distribution of abundances \cite{Ottino-Loffler20}. Quite generally, in nonlinear complex systems, one expects that the limits $t\to \infty$ and volume $\Omega \to \infty$ (or carrying capacity $\kappa \to \infty$) do not commute \cite{Qian16}: if $t \to \infty$ is taken at finite $\Omega$, then the system can explore multiple attractors and will sample states with an effective ``thermodynamic" distribution, while if $\Omega \to \infty$ is taken at finite $t$, then instead the system can be stuck in one of multiple attractors.

In both deterministic ($\omega=0$) and stochastic ($\omega>0$) forms, the GLV model is necessarily approximate because population abundances are discrete numbers, while the $n_i(t)$ in Eq.~\eqref{LV1} are continuous variables. While this difference is not expected to be critical when $n_i(t) \gg 1$, theory and simulations of Eq.~\eqref{LV1} show that abundances can vary over a wide range, with many becoming small \cite{Bunin17,Barbier18,Biroli18,Roy20}. This is particularly so in a phase with multiple attractors, for which the dynamics is chaotic \cite{Biroli18,Roy20}. In these scenarios it is important to generalize Eq.~\eqref{LV1} to a model that accounts for the fundamental discreteness of species abundances. 

Our goal here is to find a reaction network whose macroscopic behaviour yields Eq.~\eqref{LV1} in some limit. By doing so, the tools of chemical reaction network theory reviewed in section \ref{background} can be brought to bear on ecosystems. In the few-species case, the GLV model has a standard interpretation as a reaction system, for example between predator and prey, and reaction network representations have been used elsewhere in ecology \cite{Butler09,Shih14,Xue17,Constable16,Constable17}. Likewise the Doi-Peliti path-integral formulation for CRNs has been adopted for ecological models (reviewed in \cite{Dobramysl18}), but not in the many-species limit of interest here. To our knowledge there is no exposition of the conditions under which the GLV model Eq.~\eqref{LV1}, for which DMFT has been built, can be obtained from a microscopic model. The reaction network interpretation is important because it holds for an arbitrary number of individuals, and can therefore be used to disambiguate possible continuum limits. Moreover this mapping will allow us to discuss the generality of the ecology DMFT, and whether it has an analogue for true chemical reaction networks. 

\subsection{Deterministic limit}

We search for a reaction network whose dynamics coincides with that of the ecology GLV model \eqref{LV1}. As a first step, we consider the deterministic limit and leave the treatment of the noise term to the next section. In order to find the relevant set of reactions, we compare the deterministic terms in the right-hand side of \eqref{LV1} with the general reaction equation \eqref{inst1}, evaluated in the deterministic limit $\nuv=0$. We compare the chemical reaction \eqref{eq1} with the corresponding reaction rate equations \eqref{inst1} and assume, without loss of generality, the form from transition state theory \eqref{F2}, with $k_0$ obviously arbitrary. Indeed, we are free to consider a reaction and not its detailed-balanced reverse one, if necessary.  We see that the power of $n_j$ in the reaction rate equation for species $j$ and reaction $\alpha$ indicates the number of times it appears as a reactant in the reaction. 

Given that the deterministic part of the GLV model \eqref{LV1} contains a positive linear and a negative quadratic term in $n_j$, we consider the reaction pair
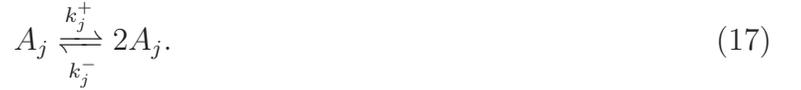
\eq{ \label{r1}
A_j \xrightleftharpoons[k^-_j]{k^+_j} 2A_j.
}
These reactions have a clear ecological interpretation as asexual reproduction in the $k^+$ direction, and competitive death in the $k^-$ direction. By identifying the kinetic factor (\refeq{F2}) with the GLV model \eqref{LV1}, we see that the pair of reactions (\refeq{r1}) appears if we take $k e^{-G_j}/(c^{eq}_j \Omega) = r_j$ and $k e^{-G_j}/(c^{eq}_j \Omega)^2 = r_j/\kappa_j$. This yields
\eqs{
\kappa_j = c^{eq}_j \Omega, \qquad k^+_j = r_j, \qquad k^-_j = (r_j \co \Omega)/\kappa_j.
} 
Therefore $\kappa_j$ is the carrying capacity of species $j$ in the absence of interactions, $r_j$ is the birth rate, and $(r_j \co \Omega)/\kappa_j$ is the rate of intra-species competition. These reactions are in detailed balance. 

We now consider the deterministic part which models the inter-species interactions, $\sum_j S_{ij} n_i n_j$. Contrary to the sum of linear and quadratic terms, which can be coupled into a pair of reactions satisfying detailed balance, the terms $S_{ij} n_i n_j$ have no reason to be paired. The reactions which give rise to the inter-species interaction term will not satisfy detailed balance, in general. In fact, consider the reactions
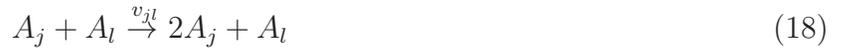
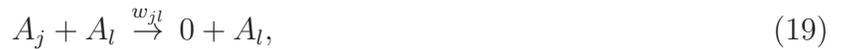
\eq{
A_j + A_l & \overset{ \;v_{jl}  }{\rightarrow} 2A_j + A_l \label{int3} \\
A_j + A_l & \overset{ \;w_{jl}\;}{\rightarrow}  0 + A_l\label{int4},
}
which give a contribution to the reaction rate equation for $j$ of the form $(v_{jl} - w_{jl}) n_j n_l$. This corresponds to a term $k e^{-G_{jl}} n_j n_l/(c^{eq}_j c^{eq}_l \Omega^2)$ in \eqref{F2} where therefore $k e^{-G_{jl}}/(c^{eq}_j c^{eq}_l \Omega^2) = v_{jl} - w_{jl}$. The inter-species interaction terms in \eqref{LV1} are recovered if $S_{jl}=w_{jl} - v_{jl}$. The rates $v_{jl}$ and $w_{jl}$ must each be nonnegative.  If $S_{jl} > 0$, then \eqref{int4} is needed, but \eqref{int3} is not, while if $S_{jl} < 0$, then \eqref{int3} is needed, but \eqref{int4} is not. Since the GLV model studied in \cite{Altieri20,Biroli18,Roy20} takes $S_{jl}$ to be random numbers with both signs, then in general both reactions are needed, and we can take
\eqs{
v_{jl} & = |S_{jl}|\Theta(-S_{jl})\\
w_{jl} & = |S_{jl}|\Theta(+S_{jl}),
}
since then $v_{jl}-w_{jl} = |S_{jl}| [ \Theta(-S_{jl}) - \Theta(+S_{jl}) ] = - S_{jl}$ as required. Here $\Theta(x)$ is the Heaviside function. 
In the ecology context, the reactions \eqref{int3} and \eqref{int4} are interpreted as cooperative growth and competitive death, respectively.

Finally, a spontaneous birth reaction is needed to recover the immigration term in \eqref{LV1}
\eq{ \label{r2}
0 & \overset{  \lambda_j  }{\rightarrow} A_j.
}

We have thus found a reaction network, defined by reactions \eqref{r1},\eqref{int3},\eqref{int4},\eqref{r2}, whose rate equations recover exactly the deterministic part of the GLV model \eqref{LV1}. This completes the first part of the mapping.

The Hamiltonian for this GLV reaction network is given by \eqref{H2}
\eq{
H = \sum_j r_j [ e^{\nu_j} - 1] \left[n_j - \frac{n_j^2 e^{-\nu_j} }{\kappa_j} \right] + \sum_{i, (j \neq i)} S_{ij} [ e^{-\nu_i} - 1 ] n_i n_j + \sum_{i, (j \neq i)} v_{ij} [e^{\hhalf\nu_i} - e^{-\hhalf\nu_i} ]^2 n_i n_j + \sum_j \lambda_j  [e^{\nu_j} -1] \label{H4}.
}

\subsection{Langevin regime}

We now consider the finite-$\omega$ regime where demographic noise plays a role. Using the reaction network \eqref{r1},\eqref{int3},\eqref{int4}, we investigate under which conditions the noise term in \eqref{LV1} is recovered. Building on the results of section \ref{langevin}, the Langevin equation for the reaction system is obtained by a small-$\nu$ expansion of the Hamiltonian \eqref{H4}
\eqs{
H & = \sum_j r_j [ \nu_j + \half \nu_j^2 ] \left[n_j - \frac{n_j^2 (1 - \nu_j) }{\kappa_j} \right] + \sum_{i, (j \neq i)} S_{ij}  [ - \nu_i  + \half \nu_i^2 ] n_i n_j + \sum_{i, (j \neq i)} v_{ij}  \nu_i^2 n_i n_j \notag \\
& \qquad + \sum_j \lambda_j [ \nu_j + \half \nu_j^2 ] + \OO(\nu^3).
}
We obtain the Langevin equation for the abundance of species $j$ using equations \eqref{Lan1}, \eqref{Lan2}, \eqref{Lan3}
\eq{  \label{Lan5}
\p_{t} n_j = r_j \left[n_j - \frac{n_j^2  }{\kappa_j} \right] - \sum_{i \neq j} S_{ji} n_i n_j + \lambda_j + \left[ r_j (n_j + n_j^2/\kappa_j) + n_j \sum_{k \neq j} [w_{jk}+v_{jk}] n_k + \lambda_j  \right]^{1/2} \xi_j,
}
which resembles that of the GLV model \eqref{LV1}, except that the noise has richer correlations.

In order to connect with recent works on the GLV model, we consider that all species have the same birth rate and carrying capacity, $r_j = r$ and $\kappa_j = \kappa$. We perform the scaling transformation
\eqs{
n_j = \kappa n_j', \quad t = t'/r, \quad w_{ij} = r w_{ij}'/\kappa, \quad v_{ij} = r v_{ij}'/\kappa, \quad \lambda_j = r \kappa \lambda_j', \quad \xi_j = r^{1/2} \xi'_j,
}
under which the Langevin equation \eqref{Lan5} becomes
\eq{ \label{LV2}
\p_{t'} n_j' = n_j'(1-n_j') - \sum_{i \neq j} S'_{ji} n_i' n_j' + \lambda_j' + \frac{1}{\sqrt{\kappa}}  \left[ n_j' + n_j'^2 + n_j' \sum_{k \neq j} [w'_{jk}+ v'_{jk}] n_k' + \lambda_j'  \right]^{1/2} \xi'_j,
}
with $\langle \xi'_j(t') \xi_k'(0) \rangle = \delta_{jk} \delta(t')$.

The GLV model Eq.~\eqref{LV1} considered in \cite{Altieri20,Altieri2022} is recovered assuming that the bracket in \eqref{LV2} can be approximated by $\alpha n_j'$, the noise strength then being given by
\eq{ \label{alpha}
2\omega^2=\alpha/\kappa.
}
To the best of our knowledge, this is the first derivation of the GLV model starting from a microscopic description. 
In the next section, we critically expose the assumptions underlying such an approximation, and detail their consequences. 

Before doing so, let us briefly discuss the difference between \eqref{LV2} and the reduced model \eqref{LV1} with simplified noise. Consider \eqref{LV2} for a generic pair of species $j$ and $k$. Suppose the abundance of $k$ is abnormally large: $n_k'\gg1$. This will affect the dynamics of species $j$ in two ways: first, and most directly, it will increase the rate of reactions in which these species are reactants, i.e. increase the magnitude of the $S'_{jk} n_j' n_k'$ term. But secondly, and more subtlely, it will increase the prefactor of the noise $\xi'_j$. Thus species $j$ becomes more {\it susceptible} to noise as a result of the fluctuation in species $k$. This susceptibility is specific, because it is multiplied by the corresponding rates $w'_{jk}$ and $v'_{jk}$. In principle, it allows for fine-tuning of states to these noise-mediated interactions. When the noise amplitude is approximated by $\alpha n_j'$, this susceptibility is lost. 

\subsection{Regime of validity of the generalized Lotka-Volterra model}

The derivation of the GLV model Eq.~\eqref{LV1} is not systematic because we need to justify (i) the approximation of the noise, and (ii) the small-$\nu$ expansion. We expose the conditions under which (i) and (ii) are justified, and therefore clarify the regime of validity of the GLV model studied recently \cite{Altieri20,Altieri2022}.

To justify (i), consider first the case of infinitesimal immigration, $\lambda=0^+$. The regime of interest is large populations, $\kappa \gg 1$. Note that the equation of motion is scaled such that $n_j'\sim 1$, typically. For such species the noise is important when $\xi'_j/\sqrt{\kappa} \sim 1$. Since the noise is scaled to be of unit magnitude, if $\kappa \gg 1$, then $\xi'_j/\sqrt{\kappa}$ is small, and noise is typically irrelevant. However, if the abundance of one species gets very small $n_j' \ll 1$, one can have $n_j' \sim \sqrt{n_j'/\kappa}$ even for large $\kappa$, i.e. when $n_j' \sim 1/\kappa$. Thus for $\kappa \gg1$ the noise term can be approximated by its value for $n_j'\ll 1$. In this case the bracket in \eqref{LV2} is approximated by 
\eqs{
n_j' \left[ 1 + \sum_{k \neq j} [w'_{jk}+ v'_{jk}] n_k' \right].
}
The rates $w'$ and $v'$ are nonnegative, so in general the second term does not cancel and must be retained. However, if we consider the large species limit $N\gg 1$ where all $w'_{jk}$ and $v'_{jk}$ are identically and independently distributed, say with mean $s/N$,
then the leading term as $N \to \infty$ will be 
\eqs{ 
n_j' \left[ 1 + 2s \overline{n}(t) + \ldots \right],
}
where $\overline{n}(t) = \ffrac{1}{N}\sum_k n_k'(t)$ is the mean rescaled abundance. For the fluctuations to be irrelevant we must assume that the $n_k'$ are not strongly correlated with $w'_{jk}-s/N$ and $v'_{jk}-s/N$. Assuming all this, we obtain a noise strength of the form \eqref{alpha} except that $\alpha = 1 + 2s \overline{n}(t)$ is time-dependent. Now, if the system fluctuates around a non-equilibrium steady state, or if the total population is explicitly fixed as in \cite{Pearce19}, then this time-dependence disappears and (i) is justified. The argument holds for finite immigration if it is subdominant in the noise, i.e. $1/\kappa \gg \lambda'$, which is equivalent to $\lambda \ll r$.

Let us try to justify (ii). As noted above, the derivation of the Langevin equation implies that $\nu_j \sim 1/\sqrt{n_j}$. A necessary condition for $\nu_j$ to take only infinitesimal values is $\kappa \to \infty$. This cannot be sufficient, because when $n_j'\sim 1/\kappa$ then clearly $\nu_j$ is not small. Thus the small-$\nu$ expansion, while justified for species of typical (large) abundance, has no clear justification if any species tends to extinction. 

To summarize, for the chemical reaction network composed of reactions \eqref{r1}, \eqref{int3}, \eqref{int4}, \eqref{r2}, if we make the following assumptions (A)
\eqs{
\text{(A1):} \quad & k^+_j = r, \quad k^-_j = (r \co \Omega)/\kappa, \quad \lambda_j = \lambda \\
\text{(A2):} \quad & \kappa \to \infty,  
}
then we obtain the Langevin equation \eqref{LV2} in rescaled variables $n_j'=n_j/\kappa, t' = t\;r$. The derivation is fully self-consistent only if all rescaled abundances are $\OO(1)$, that is, there are no rare species. In this case, the noise is typically irrelevant. 

The noise becomes important if any abundance becomes small, $n_j' \sim 1/\kappa$. In general the noise cannot be simplified beyond \eqref{LV2}, except omitting the $n_j'^2$ term. It is then necessary to make these additional assumptions
\eqs{
\text{(A3):} \quad & \text{steady state} \\
\text{(A4):} \quad & N \to \infty \\
\text{(A5):} \quad & \lambda \ll r,
}
in order to justify the noise amplitude approximation. If the GLV model with demographic noise studied in the literature \cite{Altieri20,Altieri2022} is intended to represent the macroscopic limit of the microscopic process given by the above `reactions', then the implicit assumptions which we expose in (A1-5) are necessary for its validity. Of course, it may be possible that the GLV model is justified by a different argument. For example, one could apply the renormalization group to see which terms are nontrivially irrelevant in the macroscopic limit; to our knowledge such an analysis has not been performed.

From the derivation we learn that the noise strength $\omega$ satisfies
\eqs{ 
\omega^2 = \frac{1 + 2s \overline{n} }{2\kappa} ,
}
where $\overline{n}$ is the mean dimensionless abundance, $s/N$ is the mean of rescaled reaction rates, and $\kappa$ is the carrying capacity of the species, assumed all to be equal.

The above derivation highlights several features of the GLV model. First, its main limitation is that the small-$\nu$ expansion is not justified when any species has a small population. Since demographic noise is only important when a population is small, any phenomenology present in the stochastic model which is absent in the deterministic model must be carefully scrutinized to check that it is insensitive to the correlations in the noise, which are customarily neglected. This is more than an academic point since if noise is neglected entirely, then there is a Lyapunov function, as discussed in \ref{lyapunov}. With a Lyapunov function there can be no hopping between basins, so the dynamical relevance of the multiple-attractors phase discussed in \cite{Biroli18,Roy20,Altieri20,Altieri2022} relies upon the existence of noise. 

A second feature is also highlighted, namely that when the coupling $S_{ij}$ takes both signs, it is a sum of non-negative and non-positive quantities, representing different physical processes. We will get back to this point in the discussion.

\subsection{Universality of the deterministic limit}

While the above derivation highlights subtleties in the treatment of demographic noise, the deterministic limit, obtained by taking $\kappa \to \infty$ and assuming that all rescaled abundances are $\OO(1)$,  is well-motivated. We now show that it is also {\it universal}: additional reactions can be present microscopically, without changing the GLV model. 

In particular, the reactions \eqref{int3} and \eqref{int4} are the simplest that lead to a general asymmetric interaction $S_{jl}$, but they are not unique. For example, we could add the symmetric reactions
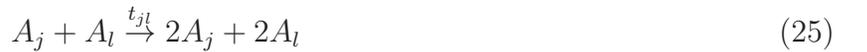
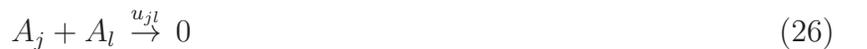
\eq{
A_j + A_l & \overset{  t_{jl}  }{\rightarrow} 2A_j + 2A_l \label{int1} \\
A_j + A_l & \overset{\;u_{jl}\;}{\rightarrow}  0 \label{int2}
}
which would also end up subsumed into $S_{jl}$. They would be distinguished at the level of the full fluctuation matrix $B_{ij}$, but under the assumptions (A1-5), and truncating the small$-\nu$ expansion at $\OO(\nu^2)$, we will obtain again \eqref{LV1}. This is a form of universality. 

Eqs. \eqref{int1} and \eqref{int2} have an ecological interpretation in terms of cooperative reproduction, and cooperative death, respectively. It is possible to add even more reactions that maintain the universal deterministic limit, but the ecological interpretation would be dubious. For example, $A_j + A_l \rightarrow m(A_j + A_l)$ also gets subsumed into the same interaction term, but does not have an obvious ecological interpretation for $m>2$.

Insensitivity to microscopic reactions is one element of universality; others have been discussed in \cite{Bunin17,Barbier18}.  

\subsection{Dynamical mean-field theory from the Doi-Peliti integral}

The DMFT for \eqref{LV1} was originally derived with the dynamical cavity technique \cite{Roy20}. Complementing this, we show here that the same DMFT can be derived from the Doi-Peliti integral. Unlike the cavity method, which works directly in the large number of species $N \to \infty$ limit, our technique is more general, and highlights the essential features necessary to obtain a DMFT.

The key advantage of treating disordered systems with dynamics is that the path integral, which is equal to one, can be directly averaged over the disorder, without replicas \cite{De-Dominicis76,DeDominicis78}. The path integral has an action $S$ given by \eqref{S1}, with the reaction network Hamiltonian \eqref{H4}. We will initially set $v_{ij}=0$ in \eqref{H4}, so that disorder only appears in the second term of the Hamiltonian via the random variables $S_{ij}$. We thus need to compute the disorder average (noted $\overline{\; \cdot \;}$) of 
\eqs{
e^{\int_t \sum_{i < j} [S_{ij} A_{ij}+S_{ji} A_{ji}]}
}
where $A_{ij}(t) = [e^{-\nu_i(t)}-1]n_i(t) n_j(t)$. Following previous works, we assume that the $\{ S_{ij} \}$ are Gaussian, with mean and variance given in \eqref{PS}.
Using standard methods, outlined in~\ref{appendixDMFT}, the disorder-averaged generating function is exactly transformed into a functional of several order parameters, each a sum over contributions from each species. The order parameters are:

\eqs{
P(t,t') & = \frac{1}{N} \sum_i e^{-\nu_i(t)} e^{-\nu_i(t')} n_i(t) n_i(t') \\
Q(t,t') & = \frac{1}{N} \sum_i e^{-\nu_i(t)} n_i(t) n_i(t') \\
C(t,t') & = \frac{1}{N} \sum_i n_i(t) n_i(t') \\
p(t) & = \frac{1}{N}\sum_i e^{-\nu_i(t)} n_i(t) \\
m(t) & =\frac{1}{N} \sum_i n_i(t),
}
and each is accompanied by a Lagrange multiplier, labelled $\Lambda_P(t,t'), \Lambda_Q(t,t'), \Lambda_C(t,t'), \lambda_p(t),$ and $\lambda_m(t)$, respectively. Although $C$ and $m$ have a clear physical meaning in terms of time-correlation and average of the species abundancies, the other order parameters do not have straightforward interpretations. Any order parameter that involves $\nu$ measures the sensitivity of abundances to fluctuations, and is therefore a type of response function. The standard, simple response function will be obtained only in the limit of small $\nu$, as explained below.

The final disorder-averaged generating function takes the form
\eq{ \label{zbar}
\overline{Z} & = \int \D[\cdots] e^{-N F} \prod_j \left( \int \D n_j \int \D \nu_j e^{-S_j} \right),
}
where 
\eqs{
\int \D[\cdots] = \int \D\Lambda_P \int \D\Lambda_Q \int \D\Lambda_C \int \D\lambda_p \int \D\lambda_m \int \D P \int \D Q \int \D C \int \D p \int \D m,
}
and
\eqs{
F & = -\frac{\sigma^2}{2} \int_{t,t'} \left[  P(t,t') C(t,t') + \gamma Q(t,t') Q(t',t) + (1+\gamma) C(t,t') [- Q(t,t')  - Q(t',t) + C(t,t') ] \right] \notag \\
& \quad + \int_{t,t'} [ \Lambda_P(t,t') P(t,t')  + \Lambda_Q(t,t') Q(t,t') + \Lambda_C(t,t') C(t,t') ] \\
& \quad+ \int_{t} [ \lambda_p(t) p(t) + \lambda_m(t) m(t) - s (p(t)-m(t)) m(t) ], 
}
and
\eqs{
S_j & = -\int_t V_j + S_{BC,j} + \int_t \nu_j \p_t n_j - \int_{t,t'} [\Lambda_P(t,t') e^{-\nu_j(t)} e^{-\nu_j(t')} + \Lambda_Q(t,t') e^{-\nu_j(t)} + \Lambda_C(t,t') ] n_j(t) n_j(t') \notag \\
& - \int_t [\lambda_p(t) e^{-\nu_j(t)} + \lambda_m(t) ] n_j(t) + \frac{s}{N} \int_t A_{jj}(t) + \frac{\sigma^2}{2N} (1+\gamma) \int_{t,t'} A_{jj}(t) A_{jj}(t'),
}
and where $V_j$ is the non-interacting part of the Hamiltonian \eqref{H4}
\eqs{
V_j(n,\nu) =  r_j [ e^{\nu} - 1] n \left[ 1 - n \frac{e^{-\nu}}{\kappa_j} \right] + \lambda_j [e^{\nu}-1].
}

For simplicity choose $S_{BC}$ so that it does not depend on $j$. Then the single-species dependence in $S_j$ comes from $r,\kappa,$ and $\lambda$. Let
\eq{
W(r,\kappa,\lambda) = \int \D n \int \D \nu \; e^{-S(r,\kappa,\lambda)}
}
be the single-species generating function, suppressing the dependence on all the Lagrange multipliers. Then we have
\eqs{
\prod_j W(r_j,\kappa_j,\lambda_j) 
& = \exp \left( \int dr \int d\kappa \int d\lambda \; \sum_j  \delta(r-r_j)\delta(\kappa-\kappa_j)\delta(\lambda-\lambda_j)\log W(r,\kappa,\lambda) \right) \notag \\
& \equiv \exp \left( N \int dr \int d\kappa \int d\lambda \;\tilde\rho(r,\kappa,\lambda) \log W(r,\kappa,\lambda) \right),
}
defining the empirical density $\tilde\rho$ of the single-species parameters. If these parameters are iid, then as $N\to\infty$, $\tilde\rho$ converges to the disorder-average density $\rho$ (more precisely the cumulative distributions converge). (At finite $N$ we can consider the simpler model where all $r_j,\kappa_j$, and $\lambda_j$ are independent of $j$.) Then we can write
 \eq{
\prod_j W(r_j,\kappa_j,\lambda_j) & \to \exp \left( N \;\overline{\log W} \right) \notag \\
& = \exp \left( N \;\overline{\log \int \D n \int \D \nu \; e^{-S} } \right)
}
so that
\eqs{
\overline{Z} & = \int \D[\cdots] \exp \left( -NF + N \;\overline{ \log \int \D n \int \D \nu \; e^{-S} } \right) 
}
This is interesting because we have already reduced the path integral from $2N$ dynamical variables to 12 dynamical variables. We did not need to assume large number of species limit $N \to \infty$, and we did not need to make any small-$\nu$ expansion. Thus this path integral still includes the effect of rare fluctuations and demographic noise to all orders in $\nu$, and could be useful to understand rare events \cite{Elgart04}. In principle, it may be subject to other analytical approximation methods that do not require large $N$.

In practice, the DMFT derivation proceeds by considering the limit $N \to \infty$, in which case the action is proportional to $N$ and the saddle-point method can be applied. The correlation and response functions are then determined by saddle-point equations
\begin{subequations} \label{SP} \eq{ 
P(t,t') & = \langle e^{-\nu(t)} e^{-\nu(t')} n(t) n(t') \rangle \label{SP1} \\
Q(t,t') & = \langle e^{-\nu(t)} n(t) n(t') \rangle, \label{SP2} \\
C(t,t') & = \langle n(t) n(t') \rangle \label{SP3} \\
p(t) & = \langle e^{-\nu(t)} n(t) \rangle, \label{SP4} \\
m(t) & = \langle n(t) \rangle, \label{SP5} \\
\Lambda_P(t,t') & = \ffrac{\sigma^2}{2} C(t,t'), \label{SP6} \\
\Lambda_Q(t,t') & = \ffrac{\sigma^2}{2} \left[ 2\gamma Q(t',t) - 2 (1+\gamma) C(t,t')] \right],\label{SP7} \\
\Lambda_C(t,t') & = \ffrac{\sigma^2}{2} \left[ P(t,t') +(1+\gamma) [- Q(t,t')  - Q(t',t) + C(t,t') ] \right], \label{SP8} \\
\lambda_p(t) & = s m(t), \label{SP9} \\
\lambda_m(t) & =s p(t) - 2 s m(t), \label{SP10}
} \end{subequations} 
where the expectation values are taken with respect to the single-species action $S_1$
\eqs{
\langle O(\nu(t),n(t')) \rangle \equiv \overline{ \left\{ \frac{\int \D n \int \D\nu \; e^{-S_1} O(\nu(t),n(t'))}{\int \D n \int \D\nu \; e^{-S_1} } \right\}}, 
}
for a general observable $O(\nu(t),n(t'))$. This reduction assumes only $N \to \infty$ with no further assumption regarding large populations. Note, however, that evaluation of the single-species expectation values requires evaluating a full 1D path integral over the effective fields $n(t)$ and $\nu(t)$. 

Since it does not require large populations, this theory generalizes the DMFT obtained previously \cite{Roy20}, and still includes rare events. However, it has a problem: we have derived this theory ignoring the $v_{ij}$ terms in \eqref{H4}, so that $S_{ij}=w_{ij}$. The distribution of $w_{ij}$ assumed in \eqref{PS} follows that of \cite{Roy20}. This is equivalent to assuming $w_{ij} = \ffrac{1}{N} [s + \sqrt{N} \sigma W_{ij} ]$ where $W_{ij} \sim \mathcal{N}(0,1)$, i.e. $\langle W_{ij} \rangle = 0, \langle W_{ij}^2 \rangle = 1$. But now we see a problem: for large $N$, $w_{ij}$ will be dominated by the random part, $w_{ij} \approx  \ffrac{\sigma}{\sqrt{N}} W_{ij}$, and therefore it will be negative for nearly half of the pairs, contradicting the fact that $w_{ij} \geq 0$ (unless $\sigma=0$). 

We face a similar problem if we begin with \eqref{H4}. Although we can let $S_{ij} = w_{ij} - v_{ij}$ take both signs, the term involving $v_{ij}$ will lead to the same problem. This result is not specific to the choice of particular reactions to compose $S_{ij}$. It is a simple consequence of the fact that all the reaction rates contribute positively to the noise, so there can be no compensation from positive and negative quantities needed for the above scaling.  

It follows that to obtain a dynamical mean-field theory for a reaction network, with the above scaling for the variance compared to the mean of the couplings, the aforementioned simplification of the noise is essential. Since this requires the small-$\nu$ expansion, to complete the derivation we now show that by expanding $S_1$ we can recover the DMFT of  \cite{Roy20}, and the closely related one of \cite{Pearce19}.

Ignoring terms of $\OO(1/N)$, we have
\eqs{
S_1 & = -\int_{t,t'} [\Lambda_P(t,t') + \Lambda_Q(t,t')+ \Lambda_C(t,t') ] n(t) n(t') - \int_t [\lambda_p(t) + \lambda_m(t) ] n(t) + \OO(\nu^3) \notag \\
& + \int_t \nu(t) [-r n(t) (1-n(t)/\kappa) - \lambda + \p_t n(t) + n(t) \lambda_p(t) ] - \int_{t,t'} \nu(t) \chi(t,t') n(t) n(t')  \notag \\
& - \half \int_{t, t'} \nu(t) \tilde K(t,t') \nu(t') + \OO(\nu^3) 
}
where
\eqs{
\chi(t,t') = -2\Lambda_P(t,t') - \Lambda_Q(t,t') 
}
and where the noise kernel is
\eqs{
\tilde K(t,t') & = \delta(t-t') \left[ r n(1+2n/\kappa) + \lambda + n \lambda_p - n(t) \int_{t''} \chi(t,t'') n(t'') \right] + 2\Lambda_P(t,t') n(t) n(t'),
}
which is coloured and multiplicative, in general. 

Now we seek a regime in which the small-$\nu$ expansion is justified. Above we saw that the correct scaling ansatz is $n \sim \kappa, t \sim 1/r$. We obtain the scaling of $\nu$ by equating the linear and quadratic terms, giving $\nu t r n \sim t^2 \nu^2 \tilde K$, i.e. $\nu \sim r^2\kappa/\tilde K$. In $\tilde K$, we have the time-local term, scaling as $(1/t) rn \sim r^2 \kappa$, and the nonlocal term, scaling as $\sigma^2 n^4 \sim \sigma^2 \kappa^4$. The nonlocal term will dominate if $\sigma^2 \kappa^4 \gg r^2 \kappa$, i.e. $\kappa^3 \gg r^2/\sigma^2$. Assuming this, we obtain $\nu \sim r^2/(\sigma^2 \kappa^3)$. Thus if $\sigma \sim 1, r \sim 1$, and $\kappa \gg 1$, then $\nu$ will indeed be small, and moreover the kernel will be dominated by its nonlocal part
\eqs{
K(t,t') & = 2\Lambda_P(t,t') n(t) n(t')
}
We now replace $\tilde K(t,t')$ by $K(t,t')$. We will check a posteriori that the time-local response term in $\tilde K$ is indeed subdominant.

Now perform a Hubbard-Stratonovich transformation
\eqs{
e^{\hhalf \int_{t,t'} K(t,t') \nu(t) \nu(t')} \propto |\det K|^{-1/2} \int \D\xi \; e^{-\hhalf \int_{t,t'} K^{-1}(t,t') \xi(t) \xi(t')} e^{\int_t \xi(t) \nu(t)}
}
so that $S_1$ is now linear in $\nu$. Integrating $\nu$ out, we obtain the nonlocal DMFT equation
\eqs{
\p_t n = [rn(1-n/\kappa) + \lambda - n \lambda_p + \xi]|_t + \int_{t'} \chi(t,t') n(t) n(t')
}
where the correlators that appear here must be determined self-consistently using \eqref{SP}. Expanding the correlators in small $\nu$ we find
\eqs{
\Lambda_P(t,t') & = \ffrac{\sigma^2}{2} C(t,t')\\
\Lambda_Q(t,t') &= \ffrac{\sigma^2}{2} \left[ -2 C(t,t') - 2 \gamma \langle  n_t n_{t'} \big( \nu_{t'} + \ldots \big) \rangle_\xi \right]
}
so that 
\eqs{
\chi(t,t') & = -\ffrac{\sigma^2}{2} \left[ 2 C(t,t') - 2 C(t,t') - 2 \gamma \langle  n_t n_{t'} \nu_{t'} \rangle_\xi +\ldots \right] \notag \\
& = \sigma^2  \gamma \langle  n_t n_{t'} \nu_{t'} \rangle_\xi +\ldots
}
For a causal solution we should have $\chi(t,t')=0$ for $t'>t$. If the average were computed using the bare propagator, this would follow from the causality rules for the Doi-Peliti path integral \cite{Tauber14}. We assume that it extends to DMFT. 
To rewrite $\chi$ in a more convenient form, note that 
\eqs{
& \int \D\xi \; e^{-\hhalf \int_{t, t'} K^{-1}(t,t') \xi(t) \xi(t')} e^{\int_t \xi(t) \nu(t)} \nu(t'') = \int \D\xi \; e^{-\hhalf \int_{t, t'} K^{-1}(t,t') \xi(t) \xi(t')}  \frac{\delta}{\delta \xi(t'')} e^{\int_t \xi(t) \nu(t)}\\ 
& \quad = -\int \D\xi \; \frac{\delta}{\delta \xi(t'')} \left[ e^{-\hhalf \int_{t, t'} K^{-1}(t,t') \xi(t) \xi(t')}  \right] e^{\int_t \xi(t) \nu(t)}\\
& \quad = \int \D\xi \; e^{-\hhalf \int_{t, t'} K^{-1}(t,t') \xi(t) \xi(t')} e^{\int_t \xi(t) \nu(t)} \int_{t} K^{-1}(t,t'') \xi(t).
}
This implies that $\nu(t'')$ has the same distribution as $\int_{t} K^{-1}(t,t'') \xi(t)$ and hence
\eqs{
\chi(t,t') = \sigma^2 \gamma \left\langle n_t n_{t'} \int_{t''} K^{-1}(t',t'') \xi(t'') \right\rangle_{\xi}.
}
If we rescale $\xi(t) = \sigma n(t) \tilde\xi(t)$, we get
\eqs{
\chi(t,t') = \sigma \gamma \left\langle n_t \int_{t''} C^{-1}(t',t'') \tilde\xi(t'') \right\rangle_{\xi}.
}
The DMFT equation becomes
\eq{ \label{lang2}
\p_t n_t = n_t \left[ 1-n_t - s\langle n_t \rangle + \sigma \tilde\xi_t  + \int_0^t dt' \chi(t,t') n_{t'} \right] + \lambda,
}
where $\langle \tilde\xi_t \tilde\xi_s \rangle = C(t,s)$. The obtained Eq.~\ref{lang2} is the same as the DMFT equation of \cite{Roy19}, where we identify $\chi$ as the response function.

We should check that the time-local response term in $\tilde K$ is subdominant. It scales as $(1/t)t \chi n^2 \sim \sigma^2 n^4 \nu \sim \sigma^2 \kappa^4 r^2 (\sigma^2\kappa^3)^{-1}  \sim r^2 \kappa$. This needs to be smaller than $\sigma^2 \kappa^4$, which requires $\kappa^3 \gg r^2/(\sigma^2)$, which certainly holds if $\sigma \sim 1$ and $r \sim 1$, as assumed.

We note in passing that the DMFT can also be derived by first expanding the Doi-Peliti action to quadratic order in $\nu$, and then performing the random average and introducing new variables.

\section{Using the mapping}
\label{mapping}

We have found a chemical reaction network (CRN) whose large-scale behavior maps to the generalized Lotka-Volterra (GLV) model recently considered in theoretical ecology. By doing so, we have discussed the underlying assumptions for the GLV model to be valid. Assuming conditions under which the reduction from a chemical reaction network (CRN) to the generalized Lotka-Volterra (GLV) model is valid, we ask how (i) this mapping may be useful in the ecology context; and (ii) how the extensive knowledge of the GLV may find application in the study of CRNs.

\subsection{From reaction networks to ecosystems}

We discuss here how the CRN theory might prove useful in the ecology context. First, as emphasized above, the existence of a Lyapunov function is a generic feature of CRNs in the deterministic limit. For the GLV model, a Lyapunov function was known \cite{Biroli18}; from the mapping we see that this is not an accident. 

Second, there is debate about the correct numerical treatment of Langevin equations with multiplicative noise \cite{Dornic05,Ceccato18}, such as the GLV model. In particular, Ref.~\cite{Altieri20} discusses how delicate the treatment of the immigration reaction in numerical simulations is. The presence of an immigration term is important to prevent extinctions. The noise, and the immigration reaction in particular, are important when a species abundance is small. In this article, we demonstrated that the reduction from a full path integral to a Langevin equation is not generally justified in this regime. Thus, the GLV CRN can be used to benchmark any treatment of the Langevin model. To this end the extensive literature on approximation and simulation of CRNs may be useful \cite{Grima11,Schnoerr14,Grima15,Wallace12}.

Third, we showed above that one can obtain a DMFT for the GLV CRN that does not require any assumption about large populations. This DMFT, while sophisticated, could be used to understand rare events in ecosystems, such as extinctions, where the Hamiltonian framework has already proved useful \cite{Kamenev08,Kamenev08a}. 

\subsection{From ecosystems to reaction networks}

We now consider, conversely, what the study of the GLV model can tell us about chemical reaction networks? A key outcome of recent works is the phase diagram of the GLV model as a function of typical interaction $s$ \cite{Bunin17,Biroli18,Roy19}, interaction fluctuation strength $\sigma$ \cite{Bunin17,Biroli18,Roy19}, interaction symmetry $\gamma$ \cite{Roy19}, and noise strength $\omega$ \cite{Altieri20, Altieri2022}. 

In the limit of high disorder (large $\sigma$), some species grow without bound and the system does not reach a steady state. This `unbounded growth' phase is however non-physical and a pathology of the model. Indeed, a large variance $\sigma$ implies the presence of strongly cooperative interactions between some of the species, that are stronger than the species' logistic saturation and lead to unbounded growth. This non-physical behavior can be cured by a saturation stronger than quadratic \cite{Roy19}.

For smaller $\sigma$ there are three dynamically stable phases, characterized by the number and type of the attractor states. The simplest is the single equilibrium phase, which holds for large enough noise strength $\omega$ \cite{Bunin17,Biroli18}. In that phase, the system relaxes toward a unique stationary state with fluctuating equilibrium dynamics, irrespective of the initial condition. When decreasing further the demographic noise $\omega$, or considering larger populations, two new phases appear, both containing multiple attractors of the dynamics \cite{Bunin17,Biroli18,Altieri20}. In these multiple-attractor phases, the equilibrium state of the system depends on the initial condition. The most remarkable phase is the one discovered at low $\omega$ in models with symmetric interactions $\gamma=1$ \cite{Altieri20}. In this region of the phase diagram, interactions play an important role, and their disordered nature gives rise to a so-called Gardner, marginally stable, phase. This phase was first discovered in the context of mean-field spin glasses \cite{mezard1987spin}, and recently shown to play an important role in the physics of jamming and amorphous materials \cite{charbonneau2014fractal,scalliet2019nature,JCPGardner}. In the Gardner phase, there is an exponential (in system size) number of equilibrium states, with a hierarchical organization in configuration space. Importantly, the states are marginally stable, i.e. poised at the edge of stability \cite{Biroli18}, and separated by arbitrarily small barriers. The presence of a fractal landscape of marginally stable minima has profound implications on the dynamics of the system. In spin and structural glasses, this gives rise to large responses (avalanches) to small perturbations, aging dynamics \cite{vincent1997slow}, as well as rejuvenation and memory effects \cite{scalliet2019rejuvenation}.

Let us suppose, for the sake of argument, that this phase structure carries over to large biochemical reaction networks; later, we will take a critical stance. The prospect of a Parisi-Gardner phase in biochemical reaction networks is tantalizing. Indeed, the notion of a landscape with many attractors has a long history in biology, going back to Waddington's notion of epigenetic landscapes \cite{Waddington14,Wang11}. While many authors have argued for a multiple-attractor landscape in biology \cite{Fang19,Qian16,Wang11,Smith11,Smith20}, explicit models of multiple attractors usually focus on the simple case of two, or few, attractors. This is the case of the Schl{\"o}gl model \cite{Schlogl72,Vellela09}, which is perhaps the simplest reaction network to support multiple equilibria, and many of its properties can be computed exactly \cite{Vellela09}. However, in a Parisi-Gardner phase, the number of equilibria scales exponentially with $N$, and one must account for the statistics of the barriers. We are not aware of any studies on chemical reaction networks where the notion of exponentially many attractor states has been entertained. Therefore it is crucial to see if the GLV DMFT, where the existence of this phase can be established, has any analogue in real CRNs. 

\section{Discussion: Towards a new universality class of DMFT}
\label{discussion}

The success of DMFT in understanding large ecosystems opens the possibility to describe the behaviour of large chemical reaction networks with similar methods. 
However, while the general goal is clear, crucial features of chemical reactions give rise to technical obstacles to overcome. Several of these have been alluded to above, and we add another below. Addressing them in turn, we argue that the current ecology DMFT is inadequate to capture generic behaviour in real CRNs. This presents an important opportunity for future work.

\subsection{Incompatibility with local detailed balance}

In extant life, many reactions are catalyzed so that the reverse reactions rarely occur, and are often neglected in modeling. Indeed, since life involves energy fluxes it must exist far from thermal equilibrium and it is expected that its current state is strongly time-asymmetric, as is now being confirmed experimentally \cite{Battle16,Tan21}. However, a complete theoretical description for chemical reactions must include each reaction as part of a pair in local detailed balance. 

We noted above that the birth-death reactions in the GLV model occur in detailed balance, but that the interactions in general do not. Let us now ask how one could modify the DMFT to include the reverse reactions. 

Consider the reaction $A_j + A_l  \overset{ \;v_{jl}  }{\rightarrow} 2A_j + A_l$ and let us now add the reverse reaction $2A_j + A_l \overset{ \;\tilde v_{jl}  }{\rightarrow} A_j + A_l  $. Local detailed balance requires that the reaction rates satisfy a constraint of the form \eqref{TST1}, i.e. $v_{jl}/\tilde v_{jl} = e^{-(\qv_\alpha-\pv_\alpha)\cdot \vec{G}}$. Without loss of generality we can adopt the form from transition state theory used for CRNs \eqref{TST2}. For the reactions with rates $v_{jl}$ and $\tilde v_{jl}$, let us denote the corresponding `activation energies' as $G^v_{jl}$. For the GLV model we need also the reactions $A_j + A_l  \overset{ \;w_{jl}  }{\rightarrow} A_l$ and $A_j + A_l \overset{ \;\tilde w_{jl}  }{\leftarrow} A_l$. Altogether these 4 reactions will contribute to the Hamiltonian
\eq{
H_{ij} = k_0 e^{-G_{ij}^v} (e^{\nu_i}-1) x_i x_j + k_0 e^{-G_{ij}^w} (e^{-\nu_i}-1) x_i x_j + k_0 e^{-G_{ij}^v} (e^{-\nu_i}-1) x_i^2 x_j + k_0 e^{-G_{ij}^w} (e^{\nu_i}-1) x_j 
}
where $x_i = n_i/e^{-G_i}$ in appropriate units. It will suffice to consider this to leading order in $\nu$:
\eq{
H_{ij} = k_0 \nu_i x_j \left[ e^{-G_{ij}^v} x_i  - e^{-G_{ij}^w} x_i - e^{-G_{ij}^v} x_i^2  + e^{-G_{ij}^w} \right] + \OO(\nu^2)
}
From here we can appreciate the problem. Earlier we combined the rates so that $v_{ij}-w_{ij}$ takes both signs and can therefore have a Gaussian distribution. This is not possible here, because the reverse reactions have different powers of $x_i$. Although for the $x_i^2 x_j$ we could add another reaction with the same factor, this is not possible with the $x_j$ reaction; physically this corresponds to the fact that extinction reactions can never be in local detailed balance. It is thus not possible to combine multiple reaction rates so that their combination has a Gaussian distribution.

In order to derive a true CRN DMFT, one must therefore deal with non-Gaussian distributions and average terms of the form
\eq{
\overline{e^{\int_t e^{-G_{ij}^v} A_{ij}}}
} 
over $G_{ij}^v$, where for example $A_{ij} = k_0 (x_i x_j (e^{\nu_i}-1) + x_j (e^{-\nu_i}-1))$. This is the moment-generating function of the random variable $y = e^{-G_{ij}^v}$, evaluated at $\int_t A_{ij}$. Further development of the theory uses the cumulant-generating function
\eq{
L(X) = \log \overline{\exp \left(y X \right)}
}
evaluated at $X= \int_t A_{ij}$. If $y$ were Gaussian, then $L$ would be a quadratic polynomial in $\int_t A_{ij}$; this leads to the two-time correlation and response quantities defined above. One might hope that an appropriate distribution would lead to a finite low-order polynomial for the non-Gaussian case. However, it is known from probability theory that no such distribution exists: there are no distributions whose cumulant generating function is a finite polynomial of order greater than two \cite{Lukacs60}. Thus $L$ will have terms of all orders in $\int_t A_{ij}$, leading to a theory with quantities depending on arbitrarily many different times. This appears completely intractable.

Summarizing this subsection, in a theory with local detailed balance, every reaction appears along with its reverse. Once these reversed reactions are included in the GLV model, it is impossible to group together multiple reactions in order for a generalized rate to have a Gaussian distribution. Non-Gaussian distributions (which are necessary to respect positivity of reaction rates) lead to an intractable DMFT. 

\subsection{Non-positivity of reaction rates}

We have seen that imposing local detailed balance leads to problems because of the positivity of reaction rates. The latter can happen even without imposing local detailed balance. Indeed, the grouping together of multiple reactions to form an effective rate $v_{ij}-w_{ij}$ that takes both signs, and therefore can be given a Gaussian distribution, is a very special feature of the GLV model. As shown above, it does not even hold beyond the leading order in $\nu$ (i.e. system size). Thus the requirement that reaction rates are positive creates an even more fundamental roadblock to construction of DMFT by path integrals than local detailed balance. A similar difficulty appears in random language models, where positivity of weights leads to the same issue, in a somewhat simpler context \cite{DeGiuli19a}. A solution to this problem might therefore have many applications.

In real ecosystems too, going beyond the Gaussian approximation is crucial to describe purely competitive systems and overcome issues related to both positive and negative interactions that arise from Gaussian distributions with finite variance.

It is worth noting that the dynamical cavity method, which can also be used to construct DFMTs, is not subject to the same limitations on non-Gaussian distributions. The cavity method and the path-integral method are complementary: while the former works directly in the limit $N\to\infty$, the latter can be used at general $N$, and can be analyzed with the sophisticated machinery of the renormalization group. In previous work \cite{Advani18}, the cavity method was used with non-Gaussian distributions for interactions in the MacArthur resource model \cite{MacArthur70,Tikhonov17}, another key model of theoretical ecology.

\subsection{Absence of conserved quantities in stoichiometry} 

Finally, the GLV model, when considered as a CRN, lacks the most conspicuous feature of real CRNs: nontrivial stoichiometry, as characterized by the stoichiometric matrix $S_{\alpha j} = q_{\alpha j} - p_{\alpha j}$. Since (non-nuclear) chemical reactions preserve the number of each element, stoichiometry gives a large number of constraints on possible reactions for a given set of species. 
However, it is easy to see that reactions of the form \eqref{r1},\eqref{int3},\eqref{int4} cannot have conserved quantities. For example, if $A + B \rightarrow A$, then $B$ cannot carry any conserved `elements'. This should be compared with a typical real reaction, such as Ca(OH)$_{2}$ + CO$_{2} \rightarrow$ CaCO$_{3}$ + H$_{2}$O, which conserves the number of calcium, oxygen, hydrogen, and carbon atoms. 
 
In some contexts, it can be permissible to consider chemical reactions that do not conserve elements. In particular, in biochemistry, reactions are often expressed up to numbers of water atoms, since water is abundant in the cellular environment. If the GLV model is to be considered as a CRN, then it too must be interpreted in a context where elements do not give stringent conservation laws.

Work on random CRNs that feature conservation laws would thus be welcome. It was already shown in \cite{De-Martino09} that adding conserved quantities (called moieties) to random CRNs strongly affects the possible growth of the network.

\subsection{Looking ahead}

Summarizing this section, we have seen that the GLV model lacks several crucial features of real CRNs. This of course does not mitigate the importance of the GLV model in ecology, but instead points to its location in the phase space of DMFT as firmly in the emergent, mesoscopic regime. Its utility there has been demonstrated: in particular, the DMFT has shown that this nonequilibrium system can exist in different phases, and universality with respect to microscopic parameters has been shown. Tackling the challenges exposed above will prove useful to describe emergent behavior in many-species chemical reaction networks.

\section{Conclusion}

Dynamical mean-field theory has recently shown its power to identify dynamical phases in a model ecosystem, the generalized Lotka-Volterra (GLV) model. In this article, we explored to what extent these recent successes can help the development of a theoretical framework for chemical reaction networks (CRN), focusing on the limit of many chemical species. 

To guide the mapping between ecology and CRN at the DMFT level, we have shown how one can construct a CRN whose infinite-size limit precisely recovers the GLV model using the Hamiltonian formalism of CRNs. We critically discussed the conditions under which this limit is obtained, and also addressed the large-but-finite-system size limit in which a Langevin description is obtained. We found that it is not possible to obtain the GLV Langevin model without quite stringent, and in general unjustified, assumptions. Then, we showed how the GLV DMFT can be obtained from the Doi-Peliti path integral, and again critically examined necessary assumptions. Finally, we discussed how the GLV-CRN mapping could be used, in both directions. Several specific features of the GLV DMFT suggest that it is not directly applicable to real microscopic CRNs: it does not have all reactions in local detailed balance, and there are no conserved moieties. These results suggest that new classes of DMFT should be developed to investigate emergent behavior in chemical reactions networks, which are relevant to describe living matter.
 
 \ack{We are grateful to Masanari Shimada for a critical reading of the manuscript. This work was funded by NSERC Discovery Grant RGPIN-2020-04762 (to E. De Giuli), a Herchel Smith Fellowship, University of Cambridge (to C. Scalliet), and a Ramon Jenkins Research Fellowship from Sidney Sussex College, Cambridge (to C. Scalliet).}
 
\appendix


\section{Doi-Peliti path integral}
\label{appendixDP}
 
\subsection{Doi formalism}

Doi \cite{Doi76} observed that the step operators in \eqref{master}, together with associated numerical factors $n_j!/(n_j-p_{\alpha j})!$, have a natural representation in terms of bosonic creation and annihilation operators, thus connecting \eqref{master} to quantum mechanics through second quantization. The vacuum, with no molecules, is denoted by $| 0 \rangle$, and by definition satisfies $\ah_i | 0 \rangle = 0$ for all $i$. A molecule of species $i$ is created by $\ahd_i$. The operators are normalized such that
\eq{
\ah_i | \nv \rangle & = n_i | \nv - \vec{1}^i \rangle, \\
\ahd_i | \nv \rangle & = | \nv + \vec{1}^i \rangle,
}
and
\eq{
[\ah_i, \ahd_j] = \delta_{ij}, \qquad [\ah_i, \ah_j] = [\ahd_i, \ahd_j] = 0.
}
From these relations one can prove that \cite{Kamenev02}
\eq{ \label{iden1}
e^{\ah_i} f(\ah_i,\ahd_i) = f(\ah_i,\ahd_i+1) e^{\ah_i},
}
for arbitrary functions $f$. 

The state of the entire system is then encoded in the state vector
\eq{
| \phi(t) \rangle = \sum_{\{\nv\}} P(\nv,t) \prod_i (\ahd_i)^{n_i} |0\rangle 
}
The master equation is then equivalent to an imaginary-time Schr\"odinger equation
\eq{ \label{schro}
\frac{\p}{\p t} |\phi(t)\rangle = \hat{H} |\phi(t)\rangle,
}
with a non-Hermitian quasi-Hamiltonian (or Liouvillian)
\eq{ \label{AH1}
\hat{H}(\vec{a}^\dagger,\vec{a}) =  \sum_\alpha \tilde k_\alpha \left[ \prod_j (\ahd_j)^{q_{\alpha j}} - \prod_j (\ahd_j)^{p_{\alpha j}} \right] \prod_i \ah_i^{p_{\alpha i}}
}
Notice that the entropic factors $n_j!/(n_j-p_{\alpha j})!$ are absent from Eqs.\eqref{schro},\eqref{H1}. 

In Eq.~\eqref{AH1}, the terms involving $(\ahd_j)^{q_{\alpha j}}$ straightforwardly represent the effect of each reaction on molecule $j$, while those involving $(\ahd_j)^{p_{\alpha j}}$ are instead present to ensure that probability normalization is preserved.

Eq.~\eqref{schro} is solved by $|\phi(t) \rangle = e^{\hat{H} t} |\phi(0)\rangle$. 
The Born rule does {\it not} apply to this construction. Instead, observables are extracted with the coherent state
\eq{
\langle \phi_0 | \equiv \langle 0 | \Pi_i \;e^{\ah_i},
}
which satisfies $\langle \phi_0 | \prod_i n_i \rangle = 1$ for any $\nv$, using Eq.~\eqref{iden1}. This implies $\langle \phi_0 | \ahd_i = \langle \phi_0|$, since $\langle \phi_0 | \ahd_i | n_i \rangle = \langle \phi_0 | n_i + 1 \rangle = 1 = \langle \phi_0 | n_i \rangle$. The expectation value of an observable $\mathcal{O}$ is given by projection onto the coherent state:
\eq{
\langle \hat{\mathcal{O}} \rangle = \langle \phi_0 | \hat{\mathcal{O}} | \phi(t) \rangle
}
For example, for the number operator $\ahd_i \ah_i$ we have
\eq{
\langle n_i(t) \rangle & = \langle \phi_0 | \ahd_i \ah_i | \phi(t) \rangle =  \langle \phi_0 | \ahd_i \ah_i \Sigma_{\{\nv\}} P(\nv,t) |\nv \rangle = \langle \phi_0 | \Sigma_{\{\nv\}} n_i P(\nv,t) |\nv \rangle = \Sigma_{\{\nv\}} n_i P(\nv,t),
}
which is indeed the expectation value of $n_i$. Let us check normalization
\eq{
\Sigma_{\{\nv\}} P(\nv,t) & = \langle \phi_0 | \phi(t) \rangle = \langle 0 | \Pi_i \;e^{\ah_i} | e^{\hat{H}(\vec{a}^\dagger,\vec{a}) t} |\phi(0)\rangle = \langle 0 | e^{\hat{H}(\vec{a}^\dagger+1,\vec{a}) t} \Pi_i \;e^{\ah_i} |\phi(0)\rangle \notag \\
& =  \langle 0 | e^{\hat{H}(1,\vec{a}) t} \Pi_i \;e^{\ah_i} |\phi(0)\rangle.
}
This will equal unity if probability is initially normalized and 
\eq{ \label{Hcon}
\hat{H}(1,\vec{a}) = 0, 
}
which is indeed satisfied by Eq.~\eqref{AH1}.

\subsection{Path integral formulation}

The path-integral representation is built just as in quantum mechanics. In the absence of sources, the partition function
\eq{
Z = \langle 0 | e^{\hat{H}(\vec{a}^\dagger+1,\vec{a}) t_f} \Pi_i \;e^{\ah_i} |\phi(0)\rangle,
}
is equal to unity $Z=1$ due to conservation of probability. This is reminiscent of the De Dominicis-Janssen field theory \cite{De-Dominicis76,Janssen76}. The path integral representation is built by dividing the time interval $[0, t_f]$ into $N \rightarrow \infty$ slices, and introducing the coherent state resolution of unity at each time slice. Following this standard procedure, one obtains the path-integral representation of the partition function \cite{Kamenev02,Peliti85,Tauber14}
\eq{
Z = \int \D\phi \int \D\phis \; e^{-S},
}
with
\eq{ \label{S11}
S = \sum_i \left[ -\phi_i(t_f) - n^0_i \log \phis_i(0) + \phi_i(0)\phis_i(0)\right] + \int_0^{t_f} dt \left[ \sum_i \phis_i \p_{t} \phi_i - H(\phisv,\phiv) \right].
}
Here the Hamiltonian becomes a function, evaluated on the fields $\phisv$ and $\phiv$ rather than $\vec{a}^\dagger$ and $\vec{a}$. This is for initial conditions $n_i(0) = n^0_i$; if instead we have Poisson-distributed initial conditions, then $n^0_i \log \phis_i(0)$ becomes $n^0_i (\phis_i(0)-1)$. 

The doubling of degrees-of-freedom in Eq.~\eqref{S11} is characteristic of dynamic problems \cite{De-Dominicis76,Janssen76,Kamenev02}. Loosely, $\phi$ can be considered as the classical part and $\phis$ as the quantum part of the concentration field \cite{Kamenev02}. However, beyond the mean-field limit, the relationship between $\phi, \phis$ and the true density field is very subtle \cite{Kamenev02,Andreanov06,Lefevre07,Smith11}. For example, the Langevin equation obtained from the Doi-Peliti action in a naive semi-classical limit has imaginary noise. This can be avoided by a Hopf-Cole transformation $\phi^*=e^\nu, \phi = n e^{-\nu}$, which makes the number operator $a^\dagger a \to \phi^*\phi = n$. 

The Doi-Peliti action \eqref{S11} is an exact rewriting of the original master equation and can be used to obtain the behaviour of arbitrary observables. It is the basis for renormalization group treatments of reaction-diffusion and related systems, when diffusion terms are added \cite{Tauber12,Tauber14}. 

\subsection{Generating function}
In order to extract observables, sources are added to $Z$ by introducing the generating function
\eq{ \label{Zzz}
Z(\zv,t) = \sum_{\{\nv\}} z_1^{n_1} z_2^{n_2} \cdots z_N^{n_N} P(\nv,t).
}
The partition function is recovered for $\zv=1$, and moments are computed by derivatives at $\zv=1$. Elgart and Kamenev \cite{Elgart04} showed that the path integral representation of the generating function Eq.~\eqref{Zzz} is similar to that of the partition function (no sources, $\zv=1$), with differences only in the boundary terms
\eq{
S[\phi,\phis;\on,\zv] =  \sum_i \left[ \phi_i(0)\phis_i(0) -z_i \phi_i(t_f) - n^0_i \log \phis_i(0) \right] + \int_0^{t_f} dt \left[ \sum_i \phis_i \p_{t} \phi_i - H(\phisv,\phiv) \right],
}
where $\on$ is the mean trajectory. As shown by Elgart and Kamenev \cite{Elgart04}, values of $\zv$ different from unity are necessary to understand rare fluctuations.

It is convenient to consider the Hopf-Cole transformation $\phi_j = n_j e^{-\nu_j}, \phis_j = e^{\nu_j}$. This is a canonical transformation with unit Jacobian. The kinetic term in the action transforms as
\eq{
\int_t \phis_j \p_t \phi_j = \int_t e^{\nu_j} e^{-\nu_j} \left[ \p_t n_j - n_j \p_t \nu_j \right] = n_j (1-\nu_j)|_{0}^{t_f} + \int_t \nu_j \p_t n_j
}
The first term can be absorbed into the boundary conditions, i.e. we now have
\eq{
& S[n,\nu;\on,\zv] =  \sum_j \left[ \phi_j(0)\phis_j(0) -z_j \phi_j(t) - n^0_j \log \phis_j(0) \right] + \sum_j n_j (1-\nu_j)|_{0}^{t_f} + \int_0^{t_f} dt \left[ \nuv \cdot \p_{t} \nv - H(\nv,\nuv) \right] \notag \\
& =  \sum_j \left[ n_j(0) -z_j n_j(t_f) e^{-\nu_j(t_f)} - n^0_j \nu_j(0) + n_j(t_f)(1-\nu_j(t_f)) - n_j(0) (1-\nu_j(0)) \right] + \int_0^{t_f} dt \left[ \nuv \cdot \p_{t} \nv - H(\nv,\nuv) \right] \notag \\
& =  \underbrace{\sum_j \left[ n_j(0) \nu_j(0) + n_j(t_f) [-z_j e^{-\nu_j(t_f)} + 1 - \nu_j(t_f) ] - n^0_j \nu_j(0) \right]}_{S_{BC}} + \int_0^{t_f} dt \left[ \nuv \cdot \p_{t} \nv - H(\nv,\nuv) \right] \label{S}
}
The particle number statistics are extracted from the generating function by contour integrals
\eq{
P(\mv,t_f) =\prod_j \oint \frac{dz_j}{2\pi i} \frac{1}{z_j^{m_j+1}} Z(\zv,t_f)
}
For example, from $P(\mv,t_f)$ the marginal distribution of the $j^{th}$ species is
\eq{
\rho_j(m_j) & = \prod_{l \neq j} \sum_{m_l \geq 0} \prod_k \oint \frac{dz_k}{2\pi i} \frac{1}{z_k^{m_k+1}} Z(\zv,t_f) \\
& = \oint \frac{dz_j}{2\pi i} \frac{1}{z_j^{m_j+1}} \prod_{k \neq j} \oint \frac{dz_k}{2\pi i} \frac{1}{z_k} \frac{1}{1- z_k^{-1}} Z(\zv,t_f),
}
where we take an integral contour $|z_k|>1$, for all $k \neq j$. Since the $z_k$-dependence in $Z(\zv,t_f)$ is restricted to a term $e^{z_k \phi_k(t_f)}$, we will have integrals of the form $(2\pi i)^{-1} \oint dz e^{z \phi}/(z-1)  = \text{Res}[e^{z \phi}/(z-1), z=1] = e^{\phi}$ so that 
\eq{
\rho_j(m_j) & = \oint \frac{dz_j}{2\pi i} \frac{1}{z_j^{m_j+1}} Z((1,\ldots,1, z_j, 1,\ldots,1),t_f) \notag \\
& = \oint \frac{dz_j}{2\pi i} \frac{1}{z_j^{m_j+1}} Z(\ov + \delta z_j \eh_j,t_f)
}
where $\ov = (1,1,\ldots,1)$ and $\eh_j = (0,\ldots,1,0,\ldots,0)$ with a 1 in the $j^{th}$ position, and $\delta z_j = z_j-1$. Therefore we need to know $Z$ in the vicinity of $\zv=\ov$.


\section{Instantons} 
\label{appendixinstanton}
As explained in the main text, we will need the stationary points of the action, \textit{i.e.} instantons.

For later use, we have $\p S/ \p z_j = - n_j(t_f) e^{-\nu_j(t_f)}$, and the first variation is
\eq{
& \delta S[n,\nu;\on,\zv] =  \sum_j \left[ \delta n_j(0) \nu_j(0) + \delta \nu_j(0) [n_j(0) - n^0_j] + \delta n_j(t_f) [-z_j e^{-\nu_j(t_f)} + 1 - \nu_j(t_f) ] \right] \notag \\
& \qquad + \sum_j \delta \nu_j(t_f) n_j(t_f) [ z_j e^{-\nu_j(t_f)} - 1 ] + \int_0^{t_f} dt \left[ \sum_i [ \delta \nu_j \p_{t} n_j +  \nu_j \p_{t} \delta n_j ] - \delta H(\nv,\nuv) \right] \notag \\
& =  \sum_j \left[ \delta n_j(0) \nu_j(0) + \delta \nu_j(0) [n_j(0) - n^0_j ] + \delta n_j(t_f) [-z_j e^{-\nu_j(t_f)} + 1 - \nu_j(t_f) ] \right] \notag \\
& \qquad + \sum_j \delta \nu_j(t_f) n_j(t_f) [ z_j e^{-\nu_j(t_f)} - 1 ] \notag \\
& \qquad +  \left[ \sum_j [ \int_0^{t_f} dt \delta \nu_j \p_{t} n_j +  \nu_j(t_f)\delta n_j(t_f) - \nu_j(0)\delta n_j(0) - \int_0^{t_f} dt \delta n_j \p_{t} \nu_j  - \int_0^{t_f} dt \delta H(\nv,\nuv) \right] \notag \\
& =  \sum_j \left[ \delta \nu_j(0) [n_j(0) - n^0_j ] + \delta n_j(t_f) [-z_j e^{-\nu_j(t_f)} + 1 ] + \delta \nu_j(t_f) n_j(t_f) [ z_j e^{-\nu_j(t_f)} - 1 ]  \right] \notag \\
& \qquad + \int_0^{t_f} dt \sum_j \left[ \delta \nu_j \p_{t} n_j - \delta n_j \p_{t} \nu_j - \frac{\p H}{\p n_j} \delta n_j   - \frac{\p H}{\p \nu_j} \delta \nu_j  \right]
}
The stationary solutions must solve
\eq{
\frac{\p n_j}{\p t} & = +\frac{\p H}{\p \nu_j} \label{Asemi1} \\
\frac{\p \nu_j}{\p t} & = -\frac{\p H}{\p n_j}  \label{Asemi2}
}
with boundary conditions
\eq{
0 & = n_j(0) - n^0_j \\
1 & = z_j e^{-\nu_j(t_f)} \\
0 & = n_j(t_f) [ z_j e^{-\nu_j(t_f)} - 1 ]
}
These are obviously degenerate and reduce to $\nu_j(t_f) = \log z_j$, $n_j(0) = n^0_j$.

Since \eqref{Asemi1},\eqref{Asemi2} take the form of Hamilton's equations, they conserve $H$:
\eq{
\frac{dH}{dt} & = \frac{\p H}{\p t} + \frac{\p H}{\p \nv} \cdot \frac{\p \nv}{\p t} + \frac{\p H}{\p \nuv} \cdot \frac{\p \nuv}{\p t} \notag \\
& = \frac{\p H}{\p t} + \frac{\p H}{\p \nv} \cdot \frac{\p H}{\p \nuv}  - \frac{\p H}{\p \nuv} \cdot \frac{\p H}{\p \nv} \notag \\
& = \frac{\p H}{\p t}
}
\\

\subsection{Mean-field instantons minimize the action} Along saddle-points, the action is
\eq{
S_c & =  -\sum_j n_j(t_f) \log z_j  + \int_0^{t_f} dt \left[ \nuv \cdot \p_{t} \nv - H(\nv,\nuv) \right] \label{S2}
}
Define the function $f(x) = 1 + e^{x}(x-1)$ and note that the convexity inequality $x - 1 \geq - e^{-x}$ implies $f(x) \geq 1 + e^x (-e^{-x}) = 0$. 
Using the form
\eq{ \label{AH2}
H_\alpha(\nv,\nuv) & = \left[ e^{\nuv \cdot (\qv_\alpha - \pv_\alpha) } - 1 \right] \underbrace{\tilde k_\alpha  \prod_i n_i^{p_{\alpha i}}}_{\equiv F_\alpha(\nv)} 
}
we have
\eqs{
\nuv \cdot \p_{t} \nv - H(\nv,\nuv) & = \nuv \cdot \frac{\p H}{\p \nuv} - H \notag \\
& = \sum_\alpha F_\alpha(\nv) \left[ \nuv \cdot (\qv_\alpha - \pv_\alpha) e^{\nuv \cdot (\qv_\alpha - \pv_\alpha) } - e^{\nuv \cdot  (\qv_\alpha - \pv_\alpha) } + 1 \right] \\
& = \sum_\alpha F_\alpha(\nv) f\left[\nuv \cdot  (\qv_\alpha - \pv_\alpha)\right] \\
& \geq 0,
}
where we are assuming that $\nuv \in \mathbb{R}^N$. Therefore
\eq{
S_c + \nv(t_f) \cdot \log \zv & \geq 0 
}
The term in the action $-\nv(t_f) \cdot \log \zv$ is used to choose observables. Aside from this term (which has no definite sign since in general $\zv$ is complex), the action is non-negative. Equality is obtained only when for each $\alpha$ and each time $t$, we have either $F_\alpha(\nv(t))=0$, or $\nuv(t) \cdot (\qv_\alpha - \pv_\alpha) = 0$. 

Let us note that mean-field solutions correspond to $\nu_j(t)=0,  \forall j,t$ and $z_j = 1 \forall j$, in which case $S_c=0$. Thus we have shown that mean-field solutions minimize the action. 

\subsection{Fluctuations around instantons} The leading fluctuations around the instanton are obtained by computing the second variation of $S$:
\eq{
\delta^2 S & =  \sum_j \left[ \delta \nu_j(0) \delta n_j(0) + \delta n_j(t_f) z_j \delta \nu_j(t_f) e^{-\nu_j(t_f)}  + \delta \nu_j(t_f) \delta n_j(t_f) [ z_j e^{-\nu_j(t_f)} - 1 ] - \delta \nu_j(t_f)^2 n_j(t_f) z_j e^{-\nu_j(t_f)} \right] \notag \\
& \quad + \int_0^{t_f} dt \left[ \delta \nuv \cdot \p_{t} \delta \nv - \delta \nv \cdot \p_{t} \delta \nuv \right] - \int_0^{t_f} dt \left[ \frac{\p^2 H}{\p \nv \p \nv} : \delta \nv \delta \nv + 2 \frac{\p^2 H}{\p \nuv \p \nv} : \delta \nuv \delta \nv + \frac{\p H}{\p \nuv \p \nuv} : \delta \nuv \delta \nuv \right] \notag \\
& =  \sum_j \left[ \delta \nu_j(0) \delta n_j(0) + \delta n_j(t_f) \delta \nu_j(t_f) - \delta \nu_j(t_f)^2 n_j(t_f) \right] \notag \\
& \quad + \int_0^{t_f} dt \left[ 2 \delta \nuv \cdot \p_{t} \delta \nv - \p_{t} [\delta \nv \cdot \delta \nuv ]\right] - \int_0^{t_f} dt \left[ \frac{\p^2 H}{\p \nv \p \nv} : \delta \nv \delta \nv + 2 \frac{\p^2 H}{\p \nuv \p \nv} : \delta \nuv \delta \nv + \frac{\p H}{\p \nuv \p \nuv} : \delta \nuv \delta \nuv \right] \notag \\
& =  \sum_j \left[ 2\delta \nu_j(0) \delta n_j(0) - \delta \nu_j(t_f)^2 n_j(t_f) \right] \notag \\
& \quad + \int_0^{t_f} dt \left[ 2 \delta \nuv \cdot \p_{t} \delta \nv \right] - \int_0^{t_f} dt \left[ \frac{\p^2 H}{\p \nv \p \nv} : \delta \nv \delta \nv + 2 \frac{\p^2 H}{\p \nuv \p \nv} : \delta \nuv \delta \nv + \frac{\p H}{\p \nuv \p \nuv} : \delta \nuv \delta \nuv \right] 
}
where in the second step we evaluated this on the instanton. We can write the bulk part of this as a Hessian operator 
\eq{
\mathcal{H} = \begin{bmatrix} -\p_{\nv}\p_{\nv} H & -\p_t \hat{\delta} -\p_{\nv}\p_{\nuv} H \\ 
\p_t \hat{\delta} -\p_{\nuv}\p_{\nv} H & -\p_{\nuv}\p_{\nuv} H \end{bmatrix},
}
acting in the space $(\nv,\nuv)$, where $(\hat{\delta})_{ij} = \delta_{ij}$.


\section{Langevin approximation}
\label{appendixlangevin}

To develop the Langevin level of approximation, we need to expand $S$ to $\OO(\nu^2)$. We have 
\eq{
S[n,\nu;\on,\zv] & =  \sum_j \left[ n_j(0) \nu_j(0) + n_j(t_f) [1-z_j + \nu_j(t_f) ( z_j - 1) - \half z_j \nu_j(t_f)^2 ] - n^0_j \nu_j(0) \right] \notag \\
& \quad+ \int_0^{t_f} dt \left[ \nuv \cdot \p_{t} \nv - \nuv \cdot \p_{\nuv} H(\nv,0) - \half \sum_{k,l} \nu_k \nu_l \p_{\nu_l} \p_{\nu_k} H(\nv,0) \right] + \OO(\nu^3)
}
Now define $B_{ij}(t) = \p_{\nu_i} \p_{\nu_j} H(\nv(t),0)$ and note that it is positive semi-definite: for any $\vec{x}$, 
\eq{
\vec{x} \cdot B \cdot \vec{x} = \sum_\alpha F_\alpha(\nv) ( \vec{x} \cdot (\qv_\alpha-\pv_\alpha))^2 \geq 0
}
so that it has a unique positive semi-definite square root.
Now perform a Hubbard-Stratonovich transformation
 \eq{
e^{\hhalf \int_t \sum_{i,j} B_{ij} \nu_i \nu_j} \propto \int \D\xi \; e^{-\hhalf \int_t \xiv^2} e^{\int_t \xiv \cdot B^{1/2} \cdot \nuv} 
}
and similarly
\eq{
e^{\hhalf z_i n_i(t_f) \nu_i(t_f)^2 }  \propto z_i^{1/2} \int d\xi^1_i \ \; e^{- \hhalf z_i (\xi^1_i)^2 }e^{(n_i(t_f))^{1/2} z_i \xi^1_i \nu_i(t_f)} 
}
so that the path integral is
\eq{ \label{AZ}
Z = \int\D n \int \D\nu \int \D\xi \; e^{-S_{BC}^0} e^{-\int_0^{t_f} dt \left[ \nuv \cdot \p_{t} \nv - \nuv \cdot \p_{\nuv} H(\nv,0) \right]} e^{-\hhalf \int_t \xiv^2} e^{\int_t \xiv \cdot B^{1/2} \cdot \nuv}. 
}
with
\eq{
S_{BC}^0 & = \sum_j \left[ \nu_j(0) \big[ n_j(0) - n^0_j \big] + \nu_j(t_f) \big[ n_j(t_f) ( z_j - 1) - (n_j(t_f))^{1/2} z_j \xi^1_j \big] \right. \notag \\
& \qquad \left. - \half \log z_j + n_j(t_f) [1-z_j ] + \hhalf z_j (\xi^1_j)^2 \right] 
}
The path integral has an action linear in $\nu$, so it can be integrated out to obtain the Langevin equation
\eq{ \label{ALan1}
0 = \p_{t} \nv - \p_{\nuv} H(\nv,0) - B^{1/2} \cdot \xiv
}
with boundary conditions
\eq{ \label{ALan2}
0 = n_j(0) - n^0_j, \qquad 0 = n_j(t_f) (z_j-1) - \sqrt{n_j(t_f)} z_j \xi^1_j.
}
Since $\xi^1 \sim \OO(1)$, from the second boundary condition we infer that $n_j(t_f) \sim z_j^2/(z_j-1)^2$. Moreover, since $\xiv^1$ has no dynamics, this boundary condition can immediately be applied to eliminate the variable, to obtain
\eq{
S_{BC}^0 & \to \sum_j \left[ n_j(t_f) [1-z_j ] + \half \frac{n_j(t_f) (z_j-1)^2}{z_j} - \half \log z_j + \log z_j + \half \log (n_j(t_f)) \right] \notag \\
& = \half \sum_j n_j(t_f) \frac{(1-z_j^2)}{z_j} + \half \sum_j \log (z_j  n_j(t_f)).
}
The logarithmic terms come from Jacobians and will be ignored in what follows, since they are of relative importance $\OO(\log n/n)$. The noise has correlations 
\eq{
\langle \xi_j(t) \rangle & = 0, \quad \langle \xi_j(t) \xi_k(t') \rangle = \delta(t-t') \delta_{jk}
} 

\subsection{Alternative boundary conditions.}

Suppose that instead of considering a generating function with $z_j$, we instead ask for the probability that $n_j(t) = m_j$. Consider this before introducing $\xi^1_j$. The $z_j$ dependence in $Z(\zv,t)$ is then fully contained in a contribution $-z_j n_j(t_f) e^{-\nu_j(t_f)}$ in the action. We need
\eq{
\oint \frac{dz_j}{2\pi i} \frac{1}{z_j^{m_j+1}} e^{z_j n_j(t_f) e^{-\nu_j(t_f)} } & = \frac{1}{m_j!}  (n_j(t_f) e^{-\nu_j(t_f)})^{m_j} \notag \\
& = \frac{1}{m_j!} n_j(t_f)^{m_j} e^{-m_j \nu_j(t_f)}
}
There is no longer any need to introduce $\xi^1_j$ and the final-time boundary condition gets modified to $0 = -n_j(t_f) + m_j$, as expected. In this case we will have
\eq{
S_{BC}^0 & \to \sum_j [n_j(t_f) - m_j \log n_j(t_f) + \log m_j! ] \notag \\
& =  \half \sum_j \log (m_j) + \OO(1)
}
where the second line uses the Stirling approximation and the boundary condition $\nv(t_f)=\mv$.

A useful variation on this is to ask that at time $t_f$ the number of species $j$ is given by $n_j(t_f) = m_j + \sqrt{m_j} \xi^1$, where $\langle \xi^1 \rangle = 0, \langle (\xi^1)^2 \rangle = 1$. Then we simply replace $m_j$ by $m_j +\sqrt{m_j} \xi^1$ in the previous result, and integrate against $e^{-\hhalf (\xi^1)^2}$. Then to obtain the modified boundary condition we have
\eq{
& \int d\nu_j e^{\nu_j(t_f) n_j(t_f)} \int d\xi^1 e^{-\hhalf (\xi^1)^2} \frac{1}{\sqrt{m_j}} e^{-(m_j +\sqrt{m_j} \xi^1) \nu_j(t_f)} \notag \\
& = \int d\xi^1 e^{-\hhalf (\xi^1)^2} \frac{1}{\sqrt{m_j}} \delta(n_j(t_f) - m_j - \sqrt{m_j} \xi^1) \notag \\
& =\frac{1}{m_j} e^{-\hhalf \left( \frac{n_j(t_f) - m_j}{\sqrt{m_j}} \right)^2}
} 
The dependence on $n_j(t_f)$ is Gaussian, which can be useful. \\
\subsection{Scaling with system size} In some applications it is convenient to work with concentrations rather than particle numbers. Moreover we noted above that $H$ carries a factor of volume. Let $H = \Omega H'$ and $\nv = \Omega \vec{c}$, $B' = \p_{\nuv} \p_{\nuv} H'$. The Langevin equation becomes $0 = \p_{t'} \vec{c} - \p_{\nuv} H'(\Omega\vec{c},0) - \Omega^{-1/2} B'{}^{1/2} \cdot \xiv$. This explicitly shows that the noise appears to be sub-dominant for large systems. 


\section{Time-reversal symmetry}
\label{appendixtrs}

For a forward-reverse reaction pair in local detailed balance, its contribution to $H$ takes the form
\eqs{ 
H^{DB}_\alpha(\nv,\nuv) & = \tilde k^+_\alpha  \left[ e^{\sum_j \nu_j (\qv_\alpha - \pv_\alpha) } - 1 \right]  \prod_i n_i^{p_{\alpha i}} + \tilde k^-_\alpha  \left[ e^{-\sum_j \nu_j (\qv_\alpha - \pv_\alpha) } - 1 \right]  \prod_i n_i^{q_{\alpha i}} \\
& = \tilde k^+_\alpha  \left[ e^{\sum_j \nu_j q_{\alpha j} } - e^{\sum_j \nu_j p_{\alpha j} } \right] \prod_i (n_i e^{-\nu_i})^{p_{\alpha i}} + \tilde k^-_\alpha  \left[ e^{\sum_j \nu_j p_{\alpha j} } - e^{\sum_j \nu_j q_{\alpha j} }  \right]  \prod_i (n_i e^{-\nu_i})^{q_{\alpha i}} \\
& =  k_0 \co \Omega e^{-G_{A_\alpha}} \left[ e^{\sum_j \nu_j q_{\alpha j} } - e^{\sum_j \nu_j p_{\alpha j} } \right] \left[  \prod_i \left(\frac{n_i e^{-\nu_i} }{c_i^{eq} \Omega}\right)^{p_{\alpha i}} -  \prod_i \left(\frac{n_i e^{-\nu_i} }{c^{eq}_i \Omega}\right)^{q_{\alpha i}} \right]
}
Now consider the (nonlocal in time) transformation
\eq{ \label{sym1}
\tilde \nu_i(t_f-t) & = -\nu_i(t) + \log (n_i(t)/(c^{eq}_i \Omega)) \\
\tilde n_i(t_f-t) & = n_i(t)
}
Then $\tilde n_i(t_f-t) e^{-\tilde \nu_i(t_f-t)} / (c^{eq}_i \Omega)= e^{\nu_i(t)}$ and $e^{\tilde \nu_i(t_f-t)} = n_i(t) e^{-\nu_i(t)}/ (c^{eq}_i \Omega)$ so that
\eq{
H^{DB}_\alpha(\tilde\nv,\tilde\nuv)|_{t_f-t} & =  k_0 \co \Omega e^{-G_{A_\alpha}} \left[ \prod_i \left( \frac{n_i(t) e^{-\nu_i(t)}}{ c^{eq}_i \Omega } \right)^{q_{\alpha i} } - \prod_i \left( \frac{n_i(t) e^{-\nu_i(t)}}{ c^{eq}_i \Omega } \right)^{p_{\alpha i} } \right] \left[  \prod_i (e^{\nu_i(t)} )^{p_{\alpha i}} -  \prod_i (e^{\nu_i(t)} )^{q_{\alpha i}} \right] \notag \\
& = H^{DB}_\alpha(\nv,\nuv)|_{t} 
}
Moreover 
\eqs{
\int_0^{t_f} d\tau \tilde \nu_j(\tau) \p_{\tau} \tilde n_j(\tau) & = - \int_0^{t_f} dt \tilde \nu_j(t_f-t) \p_{t} \tilde n_j(t_f-t)  \\
& = - \int_0^{t_f} dt  [-\nu_j(t) + \log n_j(t)/(c^{eq}_j \Omega)] \p_{t} n_j(t)  \\
& = + \int_0^{t_f} dt \nu_j(t) \p_{t} n_j(t) - \left.\left[ \log (n_j(t)/(c^{eq}_j \Omega)) n_j(t) \right]\right|_0^{t_f} + \int_0^{t_f} dt \p_{t} n_j(t) \\
& = + \int_0^{t_f} dt \nu_j(t) \p_{t} n_j(t) - n_j(t) \log \frac{n_j(t)}{e c^{eq}_j \Omega} + n_j(0) \log \frac{n_j(0)}{e c^{eq}_j \Omega}
}
Since both $H$ and the kinetic term are invariant up to boundary terms, the transformation \eqref{sym1} is indeed a symmetry of the action, when all reactions are in detailed balance. 

\section{Derivation of DMFT for GLV:}
\label{appendixDMFT}
We need to compute the disorder average (noted $\overline{\; \cdot \;}$) of 
\eq{
e^{\int_t \sum_{i < j} [S_{ij} A_{ij}+S_{ji} A_{ji}]}
}
where $A_{ij}(t) = [e^{-\nu_i(t)}-1]n_i(t) n_j(t)$. Assuming a Gaussian distribution for the $\{S_{ij}\}$ the disorder-averaged part of the action is equal to
\eq{
\overline{e^{\int_t \sum_{i < j} [S_{ij} A_{ij}+S_{ji} A_{ji}]}} = e^{\frac{s}{N} \int_t \sum_{i < j} [A_{ij}+A_{ji}]} e^{\frac{\sigma^2}{2N} \sum_{i < j} \int_{t,t'} [A_{ij}(t)A_{ij}(t')+A_{ji}(t)A_{ji}(t')+2\gamma A_{ij}(t)A_{ji}(t')]} 
}
We have
\eqs{
& \sum_{i < j} \int_{t,t'} [A_{ij}(t)A_{ij}(t')+A_{ji}(t)A_{ji}(t')] = \int_{t,t'} \left[ \sum_{i,j} A_{ij}(t)A_{ij}(t') - \sum_i A_{ii}(t) A_{ii}(t') \right] \notag \\
& =  \int_{t,t'} \left[ \sum_{i,j} [e^{-\nu_i(t)}-1][e^{-\nu_i(t')}-1]n_i(t) n_j(t)n_i(t') n_j(t') - \sum_i A_{ii}(t) A_{ii}(t') \right] \notag \\
& = \int_{t,t'} N^2 \left[ P(t,t') C(t,t') - Q(t,t') C(t,t') - Q(t',t) C(t,t') + C(t,t')^2 - \frac{1}{N^2} \sum_i A_{ii}(t) A_{ii}(t') \right],
}
where 
\eq{
P(t,t') & = \frac{1}{N} \sum_i e^{-\nu_i(t)} e^{-\nu_i(t')} n_i(t) n_i(t') \\
Q(t,t') & = \frac{1}{N} \sum_i e^{-\nu_i(t)} n_i(t) n_i(t') \\
C(t,t') & = \frac{1}{N} \sum_i n_i(t) n_i(t'),
}
and
\eqs{
\sum_{i < j} \int_{t,t'} A_{ij}(t)A_{ji}(t') & = \half \int_{t,t'} \left[ \sum_{i,j} A_{ij}(t)A_{ji}(t') - \sum_i A_{ii}(t) A_{ii}(t') \right] \\
& = \half \int_{t,t'} \left[ \sum_{i,j} [e^{-\nu_i(t)}-1][e^{-\nu_j(t')}-1]n_i(t) n_j(t)n_i(t') n_j(t') - \sum_i A_{ii}(t) A_{ii}(t') \right] \\
& = \half \int_{t,t'} N^2  \Bigg[ Q(t,t') Q(t',t) - Q(t,t') C(t,t') - Q(t',t) C(t,t') + C(t,t')^2 \\
& - \frac{1}{N^2} \sum_i A_{ii}(t) A_{ii}(t') \Bigg],
}
and
\eqs{
\sum_{i < j} [A_{ij}(t)+A_{ji}(t)] = N^2 [p(t) - m(t)] m(t) - \sum_i A_{ii}(t),
}
with
\eq{
p(t) & = \frac{1}{N}\sum_i e^{-\nu_i(t)} n_i(t) \\
m(t) & =\frac{1}{N} \sum_i n_i(t).
}
The path integral is nontrivial due to interactions between different species, represented here by nonlinear dependence of the action on the above quantities $P,Q,C,p$ and $m$. To disentangle this dependence we introduce these named quantities as new variables in the path integral, with corresponding Lagrange multipliers. Thus, we use 
\eq{
1 & \propto \int \D[\cdots] e^{-N \int_{t,t'} [\Lambda_P(t,t')(P(t,t')-\frac{1}{N} \sum_i e^{-\nu_i(t)} e^{-\nu_i(t')} n_i(t) n_i(t'))] }\notag \\
& \quad e^{-N \int_{t,t'} [ \Lambda_Q(t,t')(Q(t,t')-\frac{1}{N} \sum_i e^{-\nu_i(t)} n_i(t) n_i(t'))+\Lambda_C(t,t')(C(t,t')-\frac{1}{N} \sum_i n_i(t) n_i(t'))]} \notag \\
& \quad e^{-N \int_{t}[\lambda_p(t)(p(t)-\frac{1}{N}\sum_i e^{-\nu_i(t)} n_i(t) )+\lambda_m(t)(m(t)-\frac{1}{N} \sum_i n_i(t))] },
}
where
\eq{
\int \D[\cdots] = \int \D\Lambda_P \int \D\Lambda_Q \int \D\Lambda_C \int \D\lambda_p \int \D\lambda_m \int \D P \int \D Q \int \D C \int \D p \int \D m.
}
The disorder-averaged generating function is then
\eq{
\overline{Z} & = \int \D n\int \D \nu e^{-S_{BC}} e^{-\int_t \sum_j \nu_j \p_t n_j} \overline{e^{\int_t H}} \notag \\
& = \int \D n\int \D \nu e^{-S_{BC}} e^{-\int_t \sum_j \nu_j \p_t n_j} e^{\sum_i \int_t V_i(n_i,\nu_i)}\overline{e^{\int_t \sum_{i \neq j} S_{ij}A_{ij}}} \notag \\
& = \int \D[\cdots] e^{-N F} \prod_j \left( \int \D n_j \int \D \nu_j e^{-S_j} \right)
}
where $S_j, F,$ and $V_i$ are written out in the main text. The introduction of new quantities has allowed us to ``diagonalize'' the species dependence, such that it takes a product form, as seen in the final expression for $\overline{Z}$. This sequence of manipulations is essential to obtain a path integral that can be evaluated with the saddle-point method.

\vspace{1cm}

\bibliographystyle{iopart-num}
\bibliography{EcoRxn09nocomments}
\end{document}